\documentclass[aps,prb,twocolumn,superscriptaddress, show pacs]{revtex4-2}

\usepackage{graphicx}
\usepackage{dcolumn}
\usepackage{bm}
\usepackage{hyperref}
\usepackage{bm, amsmath}
\usepackage{txfonts}
\usepackage{xcolor}

\begin{document}

\title{Current-Induced Dynamics and Instability Pathways of Skyrmioniums in Chiral Magnets}

\author{Kaito Nakamura}
\affiliation{Department of Chemistry, School of Science \& Graduate School of Advanced Science and Engineering, Hiroshima University, 1-3-1 Kagamiyama, Higashi-Hiroshima-shi, Hiroshima 739-8526, Japan}

\author{Yuka Kotorii}
\affiliation{Mathematics Program, Graduate School of Advanced Science and Engineering, Hiroshima University, 1-7-1 Kagamiyama, Higashi-hiroshima-shi, Hiroshima, 739-8521, Japan}
\affiliation{International Institute for Sustainability with Knotted Chiral Meta Matter (WPI-SKCM2), Hiroshima University, 1-3-1 Kagamiyama, Higashi-Hiroshima, Hiroshima 739-8531, Japan} 
\affiliation{RIKEN Center for Interdisciplinary Theoretical and Mathematical Sciences (iTHEMS), RIKEN, Wako 351-0198, Japan}

\author{Andrey O. Leonov}
\thanks{Corresponding author: leonov@hiroshima-u.ac.jp}
\affiliation{Department of Chemistry, School of Science \& Graduate School of Advanced Science and Engineering, Hiroshima University, 1-3-1 Kagamiyama, Higashi-Hiroshima-shi, Hiroshima 739-8526, Japan}
\affiliation{International Institute for Sustainability with Knotted Chiral Meta Matter (WPI-SKCM2), Hiroshima University, 1-3-1 Kagamiyama, Higashi-Hiroshima, Hiroshima 739-8531, Japan}

\date{\today}

\begin{abstract}
We present a comprehensive study of current-driven dynamics, transformations, and instabilities of skyrmioniums in chiral magnetic films, considering both isolated objects and collective states forming skyrmionium-based meta-matter. Using micromagnetic simulations combined with an analytical description based on the generalized Thiele equation, we elucidate how the internal structure of skyrmioniums governs their nonequilibrium response to electric currents.

Despite carrying zero total topological charge, skyrmioniums are shown to exhibit a finite transverse velocity under applied currents. We demonstrate that this skyrmionium Hall effect originates from an imbalance between the positive and negative topological contributions of the inner skyrmion and the surrounding ring, which generally occupy different surface areas. Current-induced deformations further enhance this imbalance, leading to Hall angles that can become comparable to those of isolated skyrmions.

At higher current densities, skyrmioniums undergo distinct instability processes depending on magnetic field and uniaxial anisotropy, including elongation, collapse into an isolated skyrmion, transformation into a topologically trivial droplet, and expansion into stripe-like textures. We organize these regimes in current–field and current–anisotropy phase diagrams and resolve their microscopic pathways by monitoring the evolution of the topological charge and local rotational measures.

Beyond isolated textures, mixed skyrmion–skyrmionium lattices exhibit rich collective dynamics, including elastic transport, current-induced polymorphic transitions, soliton exchange, and stripe formation. Pulsed-current driving provides additional control, enabling access to dynamic regimes beyond continuous driving. Overall, our results establish skyrmioniums and their meta-matter assemblies as highly tunable nonequilibrium systems and reveal how current-induced transformations probe the topological energy landscape far from equilibrium.
\end{abstract}

\maketitle

\section{Introduction}
Topological solitons arise in a wide variety of physical systems as localized, particle-like configurations of continuous fields whose stability is ensured by topological protection~\cite{manton_sutcliffe,shnir,Volovik,solitons}. 
In many contexts, these nonlinear excitations do not remain isolated: mutual interactions between them can drive the formation of periodic arrangements or crystal-like superstructures. 
Such ordered assemblies represent a form of \emph{solitonic meta-matter}, in which individual solitons serve as effective building blocks, analogous to atoms in conventional materials~\cite{leonov2026metamatter}.

Among the various realizations of topological solitons and their collective phases, two-dimensional (2D) skyrmions~\cite{JETP89,Bogdanov94} have become emblematic examples in condensed-matter physics, particularly in chiral magnets (ChMs)~\cite{NT}. 
Skyrmions are classified by the second homotopy group $\pi_2(S^2)\cong\mathbb{Z}$~\cite{Faddeev,Bott}, and are assigned an integer-valued topological charge
\begin{equation}
    Q=\frac{1}{4\pi}\!\int \! \mathbf{m}\cdot\left(\partial_x\mathbf{m}\times\partial_y\mathbf{m}\right)\,dx\,dy ,
    \label{charge}
\end{equation}
where $\mathbf{m}$ is a unit vector field describing the local magnetization. 

\begin{figure*}
  \centering
  \includegraphics[width=0.99\linewidth]{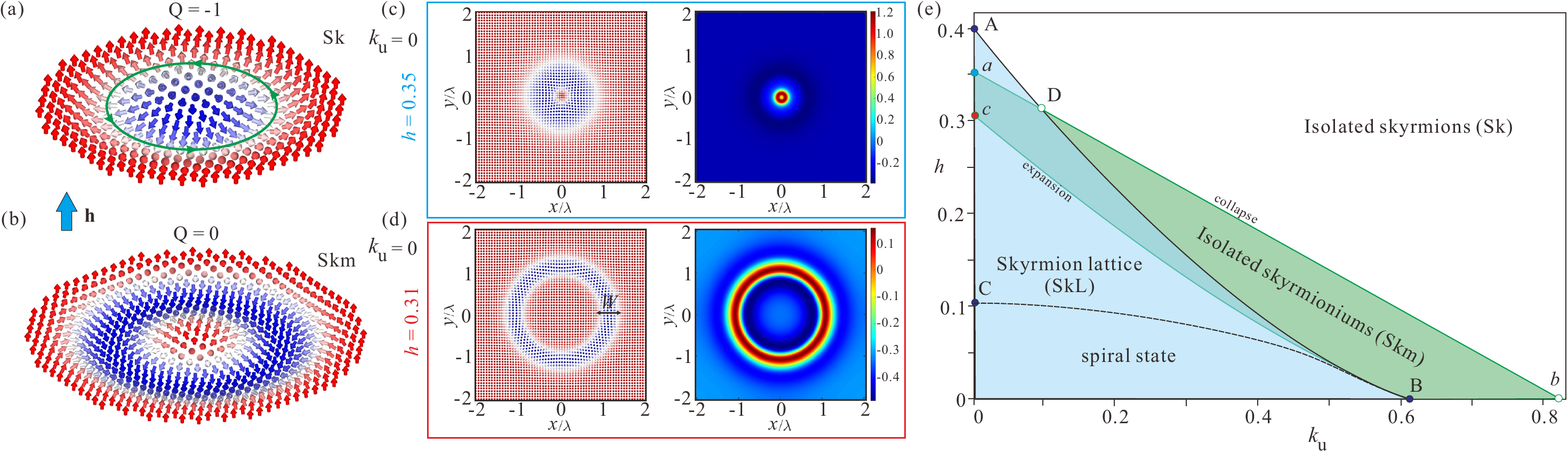}
  \caption{\label{fig01}
  (a,b) Schematic illustrations of isolated Néel skyrmions and skyrmioniums in magnetic multilayers with interfacial DMI.
  The green circle in (a) illustrates the definition of the winding number $q$.
  An isolated skyrmionium (b) is a composite texture composed of two radially nested skyrmions carrying opposite topological charges, $Q=+1$ and $Q=-1$, resulting in a vanishing total topological charge $Q=0$.
  (c,d) Structural evolution of an isolated skyrmionium within the magnetic-field interval $c$--$a$ of the phase diagram shown in (e), illustrating collapse at point~$a$ and expansion at point~$c$.
  (e) Phase diagram in the $(k_u,h)$ plane, indicating the stability regions of isolated skyrmioniums and skyrmions, as well as the skyrmion lattice and spiral phases (see text for details).
  }
\end{figure*}

In the typical skyrmion profile found in ChMs, the spins at the core
are oriented antiparallel to the surrounding uniform background and gradually
rotate to match the far-field magnetization at the outer boundary
[Fig.~\ref{fig01}(a)].
Such a configuration has $|Q|=1$ and corresponds to a magnetization texture that
wraps the entire unit sphere precisely once, thereby providing a clear geometric
interpretation of its topological character~\cite{NT,Kovalev2018}.

A skyrmion can also be characterized by a winding number $q$,  which corresponds to an element of the fundamental group $\pi_1(S^1) \cong \mathbb{Z}$.
A winding number is conventionally used to classify topological defects \cite{Selinger} but proves equally
useful for solitonic textures.
The winding number is defined along a closed loop encircling the skyrmion center
in the counter-clockwise direction (green circle in Fig.~\ref{fig01}(a)).
As one traverses this loop, the in-plane component of the magnetization rotates
by an integer multiple of $2\pi$.
For a conventional skyrmion, the magnetization completes a single
counter-clockwise rotation, yielding a winding number $q=1$. While $q$ alone does not fully determine the skyrmion topology (unlike $Q$), it provides a local and intuitive measure of rotational structure and is particularly useful when analyzing partial textures, deformations, or composite solitons.


Nowadays, magnetic multilayers provide a highly tunable 2D platform for stabilizing skyrmions. 
In these systems, interfacial Dzyaloshinskii--Moriya interaction (DMI) arises from broken inversion symmetry at the interface between heavy-metal and magnetic layers, as exemplified by PdFe/Ir(111) bilayers~\cite{Romming2013}. 
Such artificial heterostructures offer considerable flexibility—including the choice of magnetic and nonmagnetic constituents, the addition of capping layers, and the repetition of multilayer units—thereby enabling systematic control of the DMI strength, magnetic anisotropy, and resulting skyrmion characteristics.
In particular, the characteristic radius of a skyrmion~\cite{Bogdanov94,Bogdanov99} in these systems is determined by the competition between the symmetric exchange interaction and the DMI~\cite{Dz58,Moriya}. 
This balance defines a natural length scale for the skyrmion size, which can range from only a few lattice constants to several micrometers, depending on the underlying material parameters~\cite{Wiesendanger2016}.

In the past decade, magnetic skyrmions have attracted substantial interest as functional quasiparticles for next-generation spintronic technologies~\cite{Sampaio13,Tomasello14,Shigenaga}. 
Their nanoscale size~\cite{Wiesendanger2016}, topological robustness~\cite{Cortes-Ortuno}, and exceptionally low threshold for current-driven motion~\cite{Schulz,Jonietz} enable concepts such as compact and energy-efficient memory-in-logic devices~\cite{Paikaray2025}, ultrafast GHz-operation schemes, and neuromorphic computing architectures~\cite{Song2020}. 
One prominent proposal is the skyrmion racetrack memory~\cite{Wang16,Fert2013}, where information is encoded in individual skyrmions transported along confined magnetic strips. 
A major challenge for implementing such schemes, however, is the skyrmion Hall effect (SHE): under current-driven dynamics, skyrmions experience a transverse deflection that can drive them toward the track edges, ultimately leading to their annihilation~\cite{Toscano,Gobel,Yoo2017}.

Beyond conventional skyrmions, chiral magnetic materials also support a broader family of composite topological textures in which multiple skyrmion-like configurations are radially embedded within one another. 
These structures—known as $k\pi$ configurations~\cite{Bogdanov99} or target skyrmions~\cite{Leonov14}—exhibit sequential windings of the magnetization as a function of radius, giving rise to alternating skyrmionic and antiskyrmionic rings that share a common center. 
Depending on whether the number of full $\pi$ rotations is odd or even, the total topological charge alternates between $|Q|=1$ and $|Q|=0$. In particular, since a skyrmionium carries topological charge zero, the corresponding map is null-homotopic. That is, the associated magnetization configuration can be continuously deformed, through a homotopy, into a uniformly magnetized state.
Similarly, any map representing the zero element of $\pi_2(S^2)$ is homotopic to the map describing a skyrmionium, and they can be continuously deformed into one another.

A steadily growing number of experiments has demonstrated the formation and manipulation of $k\pi$ skyrmionic textures. High-symmetry confined geometries—such as magnetic nanowires~\cite{Higgins}, nanodisks~\cite{Butenko}, and nanorings~\cite{Ponsudana21}—have proven particularly suitable platforms for stabilizing these objects. 
In such nanostructures, surface-induced effects and boundary magnetization states provide additional negative energy contributions, which can energetically favor target-type configurations over competing chiral solitons~\cite{Leonov14}.
Recent work has reported target skyrmions in hybrid systems created by weakly coupling 30-nm-thick Permalloy (Ni$_{80}$Fe$_{20}$) disks of 1\,$\mu$m diameter to asymmetric (Ir 1 nm/Co 1.5 nm/Pt 1 nm)$\times$7 multilayers featuring interfacial DMI~\cite{Kent}. In parallel, off-axis electron holography has enabled direct visualization of target skyrmions in 160-nm nanodisks fabricated from the chiral magnet FeGe~\cite{Zheng}.

The skyrmionium is the simplest member of this class, corresponding to the $2\pi$ ($k\!=\!2$) case~\cite{Komineas}. 
It provides a clear example of a composite object formed from two mutually interacting skyrmionic components [Fig.~\ref{fig01}(b)]~\cite{Nakamura}. 
A skyrmionium consists of a central skyrmion with positive polarity and topological charge $Q=+1$  surrounded by a concentric ring carrying $Q=-1$. 
The characteristic radii of these two constituent textures are not independent: the size of the inner skyrmion influences the width and position of the outer ring, and vice versa. 
In the limit of a large inner skyrmion, the entire configuration squeezes  toward a thin circular domain wall (Fig. \ref{fig01} (d)), whereas a very small skyrmion core forces the outer ring to expand in order to maintain the required topological winding and local stability [Fig.~\ref{fig01}(c)].

Skyrmionium was observed experimentally in 2013 \cite{Finazzi}, its creation, detection, and motion driven by spin-polarized current have been investigated in Ref. \cite{Qiu2024,Komineas}. 
Skyrmioniums have also been realized in extended thin-film geometries without the need for lateral confinement \cite{jefremovas2024}. Examples include ferromagnetic films exchange-coupled to magnetic topological insulators~\cite{Zhang}, frustrated Kagome magnets such as Fe$_3$Sn$_2$~\cite{Yang23}, and exfoliated flakes of the van der Waals compound Fe$_{3-x}$GeTe$_2$~\cite{Pwoalla23}. The stability of skyrmioniums has been systematically investigated from a theoretical perspective as well. Energy barriers connecting Skm and ordinary skyrmion states were computed using the geodesic nudged elastic band approach in Ref.~\cite{Hagemeister}. More recent micromagnetic studies have analyzed the thermal decay of skyrmioniums and their possible topological transformations~\cite{Jiang24}, uncovering several distinct annihilation pathways and conversions into conventional skyrmions.

Such an  idea of “a skyrmion within a skyrmion” is particularly attractive as a way to overcome certain limitations associated with conventional skyrmions. For example, unlike ordinary skyrmions, skyrmioniums possess a vanishing net topological charge ($Q \!\approx\! 0$), as their magnetization field first wraps and then unwraps the unit sphere when the spatial coordinates $(x,y)$ sweep the plane, resulting in an almost complete cancellation of the topological winding. This property is known to suppress the transverse deflection of the structure, enabling more controlled, rectilinear motion in racetracks. In addition to this topological advantage, skyrmioniums are shown to exhibit enhanced mobility~\cite{Kolesnikov18,Wang20} and can reach higher velocities in comparison with  skyrmions \cite{Zhang2016,Gobel2019}. 
Thus, implementing these new magnetic configurations alongside skyrmions is crucial for logic devices.
As an example, a reconfigurable nano-magnetic logic devices based entirely on skyrmionium
dynamics were recently proposed in Refs.~\cite{Paikaray2025,Vilca2025}, where the logic
operation can be switched between OR and AND gates through voltage-controlled
modulation of the magnetic parameters.

In the present paper, we revisit the current-driven dynamics of isolated
skyrmioniums in quasi-two-dimensional chiral magnets with easy-axis
anisotropy. {\color{black}While previous studies commonly assume that skyrmioniums move strictly parallel to the driving current due to the cancellation of their net
topological charge, several subtle aspects of their dynamics remain
insufficiently understood. In particular, the origin and possible magnitude of
a residual skyrmionium Hall response have not been clarified.}

Here we demonstrate that a small but finite Hall effect may arise from the
discrete internal structure of a skyrmionium. Even a slight mismatch between
the effective topological charges of its two constituent skyrmions leads to a
nonvanishing gyrotropic response and hence to a lateral deflection during
motion. Although this effect is typically weak and often neglected in
continuum descriptions, it becomes increasingly relevant for skyrmioniums of
small size and in regimes where the magnetization varies rapidly in space.
Remarkably, the skyrmionium Hall effect is strongly enhanced near the collapse
instability (e.g., point~$a$ in Fig.~\ref{fig01}(e)), where it becomes
comparable in magnitude to the Hall response of conventional skyrmions.

We further investigate the stability of isolated skyrmioniums under large
current densities and identify several dynamical instability mechanisms.
In this regime, skyrmioniums act as sensitive probes of energetically more
stable magnetic phases and reveal a rich variety of topological transitions.
{\color{black}In addition to collapsing into an ordinary skyrmion, a skyrmionium may transform into a droplet state, followed by further elongation into extended
phases such as domain structures with chiral domain walls or arrays of kinks.
This transformation establishes a direct connection between a localized
topologically trivial object ($Q=0$) and the homogeneous magnetic state with
the same topological charge.}
Near the boundary of the spiral phase, skyrmioniums tend to elongate without bound while still preserving their topological identity.
{\color{black}Importantly, such instability mechanisms cannot be captured within the framework of the Thiele equation. The Thiele approach assumes a rigid
solitonic texture whose internal structure remains unchanged during motion.
In contrast, the transformations observed here involve substantial
rearrangements of the magnetization field, including changes in the internal
structure and topology of the skyrmionium. These processes therefore require
a full micromagnetic description beyond the collective-coordinate treatment.}

For comparison, we also analyze the impact of alternating (ac) currents on
skyrmionium dynamics and show that the resulting dynamical characteristics
correspond to time-averaged counterparts of the dc-driven motion. 

Finally, we extend the recently introduced concept of skyrmionium meta-matter—mixed lattices of skyrmions and skyrmioniums introduced in
Ref.~\cite{leonov2026metamatter}—by investigating their current-driven dynamics.
Such composite crystals exhibit a variety of collective transport regimes
depending on the lattice stoichiometry and the underlying plane group. In
certain configurations, skyrmions are forced to hop between neighboring
potential wells formed by the surrounding skyrmioniums, whereas in other
cases their motion entrains the entire lattice, causing the crystal to
translate as a single composite soliton. As a representative example, we
demonstrate current-induced transitions between different polymorphs of
skyrmionium meta-matter.

\section{Phenomenological theory of skyrmioniums in two-dimensional helimagnets \label{section:phenomenology}}

\subsection{Micromagnetic model\label{sect:model}}

The energetics of a monolayer system in which interfacial DMI is induced can be described by an energy density composed of exchange, DMI, Zeeman, and uniaxial anisotropy contributions:
\begin{equation}
w(\mathbf{m}) = \sum_{i,j} (\partial_i m_j)^2 + w_{\mathrm{DMI}} - \mathbf{m}\!\cdot\!\mathbf{h} - k_u m_z^2 .
\label{functional}
\end{equation}
Throughout this work, spatial coordinates $\mathbf{x}$ are expressed in units of the intrinsic length scale of the modulated states,
$L_D = A/D$.
Here $A>0$ denotes the exchange stiffness, $D$ the Dzyaloshinskii constant, and $k_u = K_u M^2 A / D^2$ the corresponding dimensionless uniaxial anisotropy parameter. We focus exclusively on the easy-axis case, $k_u>0$.

For systems with C$_{nv}$ symmetry, the interfacial DMI contribution takes the standard form
\begin{equation}
w_{\mathrm{DMI}} = 
m_x \partial_x m_z - m_z \partial_x m_x 
+ m_y \partial_y m_z - m_z \partial_y m_y .
\label{DMI}
\end{equation}

For quantitative comparisons with experiments, the dimensionless model (\ref{functional}) can be readily mapped to physical units.
As an illustrative example, we consider Co/Pt multilayers with typical parameters
$D = 3$ mJ/m$^2$, 
$M_s = 580$ kA/m, 
and $A = 15\times 10^{-12}$ J/m.
These values yield a characteristic length $L_D = 5$ nm, which is the unit of length used in the figures.

Alternatively, it is often convenient to employ the helical period as the unit of length
\begin{equation}
\lambda = 4\pi L_D ,
\label{lambda}
\end{equation}
which corresponds to the spiral wavelength at zero magnetic field 
(e.g., $\lambda\!\approx\!18$~nm in MnSi and $\approx\!60$~nm in Cu$_2$OSeO$_3$~\cite{Seki2012,Crisanti}).

As a model system for skyrmionium dynamics, we consider a thin ferromagnetic film lying in the $xy$ plane and impose periodic boundary conditions along both in-plane directions.
The external magnetic field is introduced in dimensionless form as
$\mathbf{h}=\mathbf{H}/H_0$,
where $H_0 = D^2/(A|\mathbf{M}|)$.
The magnetization is described by the unit vector field
$\mathbf{m}(x,y)=\mathbf{M}/|\mathbf{M}|$.

%

To minimize the functional~\eqref{functional} and obtain the static solution of the functional (\ref{functional}), we employ the \textsc{mumax3} micromagnetic package (version 3.10), which solves the Landau--Lifshitz--Gilbert (LLG)  on a finite-difference grid~\cite{mumax3}. 
To ensure that the solutions are robust and not artefacts of a particular numerical scheme, we additionally verified the results using independent in-house codes based on simulated annealing and single-step Monte Carlo relaxation with the Metropolis algorithm.  
These routines are described in detail in Ref.~\cite{metamorphoses} and are not repeated here.

\begin{figure*}
  \centering
  \includegraphics[width=0.9\linewidth]{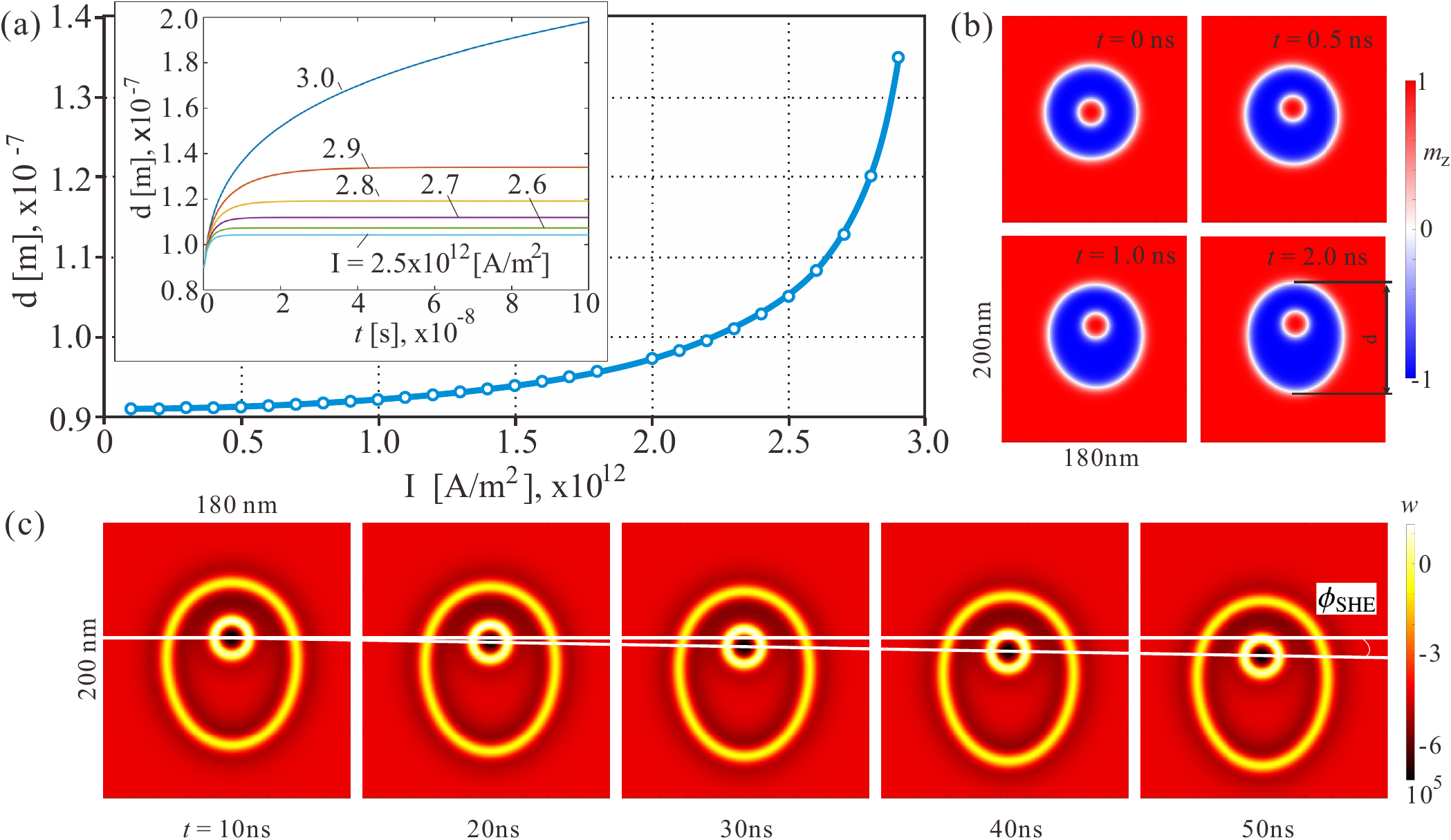}
  \caption{\label{fig02}
  (a) Current-induced elongation of an isolated skyrmionium at $k_u=0.7$, $h=0$,
  shown as the dependence of the equilibrium size $d$ on the applied current density.
  As illustrated in the inset, within $\approx 2\,\mathrm{ns}$ the skyrmionium reaches
  its steady-state size $d$, as also visualized by the color maps of the $m_z$
  component of the magnetization in (b).
  (c) After a relatively long propagation time, a small but finite SHE,
  characterized by the Hall angle $\phi_{\mathrm{SHE}}$, becomes discernible, as shown
  by the color maps of the energy density distribution at equally spaced time
  intervals. {\color{black}The white lines are guides to the eye illustrating the displacement   of the skyrmionium center; they do not represent the actual value of the Hall   angle, which remains very small.}
  }
\end{figure*}

\subsection{Phase diagram of states}

The phase diagram of equilibrium solutions of the micromagnetic model~\eqref{functional} in the $(h,k_u)$ control-parameter plane has been reported previously~\cite{Mukai} and is reproduced here in Fig.~\ref{fig01}(e).

One-dimensional (1D) spiral states occupy the curvilinear region labeled $c$--$B$--$0$. 
Along the boundary $c$--$B$, the spiral wavelength diverges, signaling the emergence of repulsive isolated kinks whose eigenenergy is positive relative to the homogeneous state. 
The corresponding critical control parameters at this boundary are $h(c)=0.308$ and $k_u(B)=0.617$ in dimensionless units, which translate to $\mu_0 H(c)=0.3186\,\mathrm{T}$ and $K_u(B)\approx3.7\times10^{5}\,\mathrm{J/m^3}$ in physical units.

For magnetic fields below this boundary, isolated 1D kinks acquire negative eigenenergy and condense into an equilibrium periodic array with a finite inter-kink spacing.

2D skyrmion lattices (SkLs) are stable in the region $A$--$B$--$0$ and exhibit closely related phenomenology. The corresponding critical control parameter at this boundary is $h(A)=0.401$  in dimensionless units, which translates to $\mu_0 H(c)=0.414 \,\mathrm{T}$  in physical units.
Upon approaching the boundary $A$--$B$ from above, 
the eigenenergy of an isolated skyrmion becomes negative; as a result, isolated skyrmions condense into an extended SkL.
Throughout the parameter range considered, the SkL stability domain lies above that of the spiral state, and a first-order phase boundary separating these two modulated phases runs along $C$--$B$. $h(C)=0.116, \mu_0H(C)=0.12$T. 
Importantly, the tricritical point reported in earlier studies~\cite{Bogdanov94,Bogdanov99,Butenko2} does not appear in the present phase diagram. 
As demonstrated in Ref.~\cite{nanomaterials2024}, slightly below the $A$--$B$ line the hexagonal SkL undergoes a continuous transformation into a square SkL, thereby expanding the overall SkL stability region and eliminating the previously inferred critical point.

Isolated skyrmioniums form a distinct branch of solutions confined to the narrow region $a$--$b$--$B$--$c$ of the phase diagram~\cite{Nakamura}. 
The corresponding critical control parameters at this boundary are $h(a)=0.352$ and $k_u(b)=0.82$ in dimensionless units, which translate to $\mu_0 H(a)=0.365\,\mathrm{T}$ and $K_u(b) \approx 4.9\times10^{5}\,\mathrm{J/m^3}$ in physical units.

Along the boundary $a$--$b$, a skyrmionium collapses into a conventional skyrmion (Fig. \ref{fig01} (c)), whereas along $c$--$B$ it continuously evolves into a circular domain wall whose radius diverges [Fig.~\ref{fig01}(d)] but the width $W$ remains finite. 

This phase-diagram topology implies an intrinsic instability of skyrmionium lattices within the present model (see for details Ref. \cite{leonov2026metamatter}). 
Specifically, the eigenenergy of an expanding isolated skyrmionium never becomes negative with respect to the homogeneous state, thereby preventing condensation into an extended skyrmionium lattice (SkmL). 
This behavior contrasts with that of spiral and skyrmion solutions, for which the eigenenergy of isolated solitons eventually crosses into negative values, enabling lattice formation while maintaining a finite soliton size. 
Consequently, within this geometry isolated skyrmioniums cannot serve as stable quasi-atomic building blocks of a periodic SkmL. 

In practice, lowering the applied magnetic field below the skyrmionium metastability window renders isolated skyrmioniums elliptically unstable, causing them to expand and reconstruct the equilibrium spiral state with its characteristic period. 
Below the critical field $h(c\text{--}B)$, the rotational DMI contribution associated with the narrow ring-like domain wall of a skyrmionium dominates, driving the eigenenergy negative and inducing the expansion of the wall—whether flat or circular—until it fills the plane, in close analogy with the formation of a simple spiral.

\section{Current-driven dynamics of isolated skyrmioniums}

\subsection{Micromagnetic simulations}

The current-driven dynamics of isolated Skms within the metastability region $a$--$b$--$B$--$c$ of the phase diagram [Fig.~\ref{fig01}(e)] was simulated using \textsc{MuMax3}, which numerically solves the LLG equation augmented by current-induced spin-transfer torque (STT) terms~\cite{mumax3,Joos}:
\begin{equation}
\frac{\partial \mathbf{m}}{\partial t}
=
-\gamma_{0}\,\mathbf{m}\times \mathbf{H}_{\mathrm{eff}}
+ \alpha\,\mathbf{m}\times \frac{\partial \mathbf{m}}{\partial t}
- (\mathbf{u}\!\cdot\!\nabla)\mathbf{m}
+ \xi\,\mathbf{m}\times\!\bigl[(\mathbf{u}\!\cdot\!\nabla)\mathbf{m}\bigr].
\label{LLG-current}
\end{equation}
Here, $\gamma_{0}$ denotes the (positive) gyromagnetic ratio; for electrons one typically uses
$\gamma_{0}=\gamma\mu_{0}\approx2.211\times10^{5}\,\mathrm{m/(A\,s)}$.
The effective magnetic field is defined as the functional derivative of the total micromagnetic energy,
\begin{equation}
\mathbf{H}_{\mathrm{eff}}
=
-\frac{1}{\mu_0 |\mathbf{M}|}\,
\frac{\delta w}{\delta \mathbf{m}} .
\end{equation}
where $w$ is given by the energy functional~\eqref{functional}.
The Gilbert damping constant is fixed to $\alpha=0.1$ in all simulations.

The term $(\mathbf{u}\!\cdot\!\nabla)\mathbf{m}$ describes the adiabatic spin–transfer torque.
The spin-drift velocity is
\begin{equation}
\mathbf{u}
=\frac{P\mu_{B}}{eM_{s}}\,\mathbf{j},
\end{equation}
where $P$ is the spin polarization ($P=0.4$), $\mu_{B}=9.274\times 10^{-24}\,\mathrm{J/T}$ is the
Bohr magneton, $e=1.602\times10^{-19}\,\mathrm{C}$ is the elementary charge, $M_{s}$ is the
saturation magnetization, and $\mathbf{j}$ is the applied current density which has only $x$-component for simplicity, i.e., $u_y=0$. The final term in Eq.~\eqref{LLG-current},
$\xi\,\mathbf{m}\times[(\mathbf{u}\cdot\nabla)\mathbf{m}]$, represents the
corresponding non-adiabatic spin–transfer torque.
The dimensionless parameter $\xi$ controls the degree of non-adiabaticity of the torque and is
typically in the range $0.01$–$0.5$; in our simulations we use $\xi=0.05$.

\textcolor{black}{We note that the present model includes only STT as the driving mechanism. In ultrathin ferromagnet/heavy-metal systems with interfacial Dzyaloshinskii--Moriya interaction, spin--orbit torques (SOT) are often dominant and can significantly influence the current-driven dynamics. In this work, however, we intentionally restrict ourselves to STT in order to isolate the fundamental instability mechanisms of skyrmioniums within a minimal and controlled framework. The inclusion of SOT would introduce additional parameters and material-specific effects, which is beyond the scope of the present study. A systematic investigation of SOT-driven dynamics will be addressed in future work.}

As an instructive example, we first simulate the spin-polarized current-driven motion of an isolated skyrmionium at the point $k_{u}=0.7$, $h=0$ of the phase diagram, which corresponds---using the Co/Pt material parameters introduced above---to a physical anisotropy
$K_u \approx 4.2\times10^{5}\,\mathrm{J/m^{3}}$.

As shown in Fig.~\ref{fig02}(a), over the wide range of applied current densities the skyrmionium exhibits the same qualitative response.
Within a short time interval $\Delta t \approx 2\,\mathrm{ns}$ (see the inset of Fig.~\ref{fig02}(a), which shows the time evolution of the lateral size $d(t)$ for a range of current densities), the texture undergoes a pronounced deformation: the inner skyrmion is displaced upward, while the outer ring is dragged downward \cite{Zhang2016} [Fig.~\ref{fig02}(b)].
In other words, the skyrmionium rapidly loses its initial rotational symmetry.
This deformation can be naturally understood as a consequence of the opposite SHEs acting on the inner and outer parts of the skyrmionium, which carry opposite partial topological charges \cite{Zhang2016}.
After the transient interval $\Delta t$, the skyrmionium no longer undergoes any
appreciable shape deformation. In practice, it follows an almost perfectly
linear trajectory aligned with the current direction, exhibiting no visible
lateral displacement. Figure~\ref{fig02}(a) shows such an equilibrium skyrmionium size.

Nevertheless, a very small residual Hall deflection [Fig.~\ref{fig02}(c)]
persists and becomes discernible only after sufficiently long propagation
times. In the following section, we address the question of whether the Skm Hall effect
is merely a numerical artefact or whether it is supported by a genuine physical
mechanism.

\subsection{Thiele equation and the origin of the skyrmionium Hall effect}
To elucidate the current-driven dynamics of a skyrmionium, we adopt the
collective-coordinate (Thiele) approach~\cite{Thiele}, following a line of analysis
similar to that of Ref.~\cite{Zhang2016}.
Within this framework, the magnetization texture is treated as a rigid
quasiparticle---i.e., after the initial transient interval $\Delta t$, during
which the skyrmionium deforms to its equilibrium shape (see inset of
Fig.~\ref{fig02}(a))---that translates without internal deformation at a
constant velocity
$\mathbf{v}$,
\[
\mathbf{m}(\mathbf{r},t)=\mathbf{m}_0(\mathbf{r}-\mathbf{v}t).
\]
Projecting the Landau--Lifshitz--Gilbert equation onto the translational mode then
yields the generalized Thiele equation
\begin{equation}
    -\,\mathbf{G} \times (\mathbf{v} - \mathbf{u})
    + \mathbf{D} \cdot (\alpha \mathbf{v} - \xi \mathbf{u})
    + \mathbf{F} = 0,
\label{Thiele}
\end{equation}
where $\mathbf{G}$ is the gyrocoupling vector, $\mathbf{D}$ is the dissipative tensor.  

Equation~(\ref{Thiele}) comprises three physically distinct contributions:
(i) the gyroscopic (Magnus) force $-\mathbf{G}\times(\mathbf{v}-\mathbf{u})$, which
induces a transverse motion and is responsible for the Hall-like deflection of
chiral spin textures;
(ii) the dissipative drag force encoded in the tensor $\mathbf{D}$, which governs
the longitudinal response and energy dissipation; and
(iii) the conservative force $\mathbf{F}= -\partial W/\partial \mathbf{X}$, which may originate from external potentials, sample boundaries, or other inhomogeneities. In the present geometry, translational invariance of the system implies $\mathbf{F}=0$.

The gyrovector is defined as
\[
\mathbf{G}
= \int \mathrm{d}^2 r\;
\mathbf{m} \cdot
\left(
\frac{\partial \mathbf{m}}{\partial x}
\times
\frac{\partial \mathbf{m}}{\partial y}
\right)
\,\hat{\mathbf{z}}
= (0,0,G_z)
= (0,0,4\pi Q),
\]
where $Q$ is the topological charge introduced in Eq.~(\ref{charge}).

The symmetric dissipative tensor is given by
\[
D_{ij}
= \int \mathrm{d}^2 r\;
\frac{\partial \mathbf{m}}{\partial x_i}
\cdot
\frac{\partial \mathbf{m}}{\partial x_j}.
\]
All definitions introduced above are consistent with the numerical implementation
of the Thiele formalism in \textsc{MuMax3}.

Introducing $\mathbf{v}=(v_x,v_y)$ and $\mathbf{u}=(u_x,u_y)$,
Eq.~(\ref{Thiele}) can be written as a linear $2\times2$ system,
\[
\underbrace{
\begin{pmatrix}
\alpha D_{xx} & -G_z + \alpha D_{xy} \\
G_z + \alpha D_{xy} & \alpha D_{yy}
\end{pmatrix}
}_{\mathbf{M}}
\begin{pmatrix}
v_x \\[4pt]
v_y
\end{pmatrix}
=
\underbrace{
\begin{pmatrix}
\xi \left( D_{xx} u_x + D_{xy} u_y \right) - G_z u_y \\
\xi \left( D_{xy} u_x + D_{yy} u_y \right) + G_z u_x
\end{pmatrix}
}_{\mathbf{b}} .
\]
Solving the matrix equation $\mathbf{M}\mathbf{v}=\mathbf{b}$ yields explicit
expressions for the velocity components:
\begin{align}
v_x &= \frac{1}{\det\mathbf{M}}
\Bigl[
\alpha D_{yy}\Bigl(
   -G_z u_y
   + \xi D_{xx} u_x
   + \xi D_{xy} u_y
\Bigr)
\nonumber\\
&\quad
-\left(\alpha D_{xy}-G_z\right)
\Bigl(
   G_z u_x
   + \xi D_{xy} u_x
   + \xi D_{yy} u_y
\Bigr)
\Bigr],
\nonumber\\[0.4em]
v_y &= \frac{1}{\det\mathbf{M}}
\Bigl[
-\left(\alpha D_{xy}+G_z\right)
\Bigl(
   -G_z u_y
   + \xi D_{xx} u_x
   + \xi D_{xy} u_y
\Bigr)
\nonumber\\
&\quad
+\alpha D_{xx}
\Bigl(
   G_z u_x
   + \xi D_{xy} u_x
   + \xi D_{yy} u_y
\Bigr)
\Bigr].
\label{vy_tc}
\end{align}
where $\det\mathbf{M}
=
\alpha^2 D_{xx}D_{yy}
-\left(\alpha D_{xy}-G_z\right)\left(\alpha D_{xy}+G_z\right)
=
\alpha^2\left(D_{xx}D_{yy}-D_{xy}^2\right)+G_z^2$.

Setting $G_z=0$, the velocity components simplify considerably,
\begin{equation}
v_x = \frac{\xi}{\alpha}\,u_x,
\qquad
v_y = \frac{\xi}{\alpha}\,u_y =0,
\label{ideal}
\end{equation}
which demonstrates the absence of any transverse (Hall) motion in the
topologically trivial case $G_z=0$.
Consequently, the skyrmionium velocity is strictly collinear with the driving
current, despite the elongated skyrmionium shape with $D_{xx}\neq D_{yy}$.

\begin{figure}
  \centering
  \includegraphics[width=0.99\linewidth]{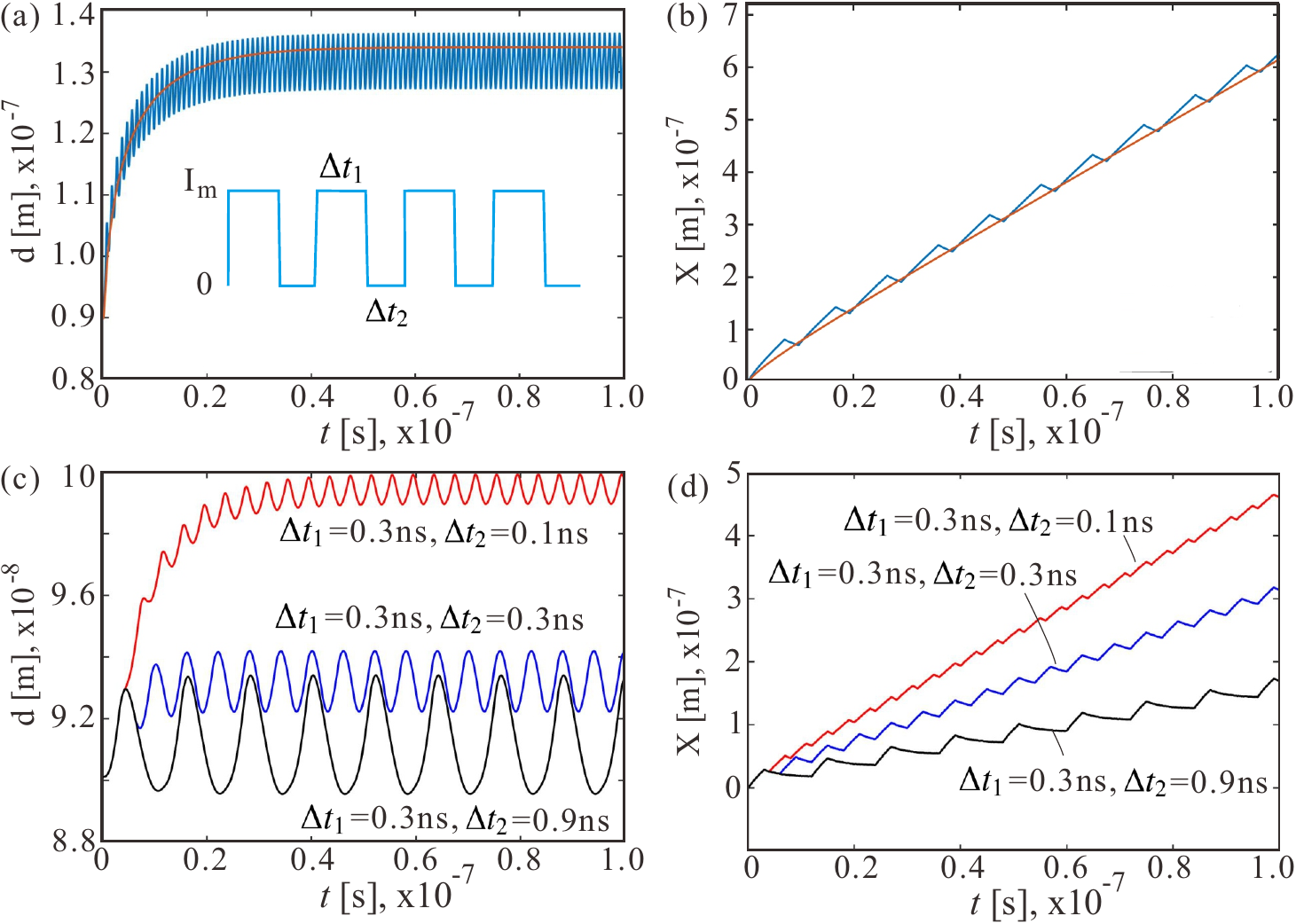}
  \caption{\label{fig08}
Current-driven dynamics of an isolated skyrmionium under pulsed and continuous driving. 
(a) Temporal evolution of the skyrmionium size for a pulsed current with peak amplitude 
$I_m = 4.0 \times 10^{12}\,\mathrm{A/m^{2}}$ applied during $\Delta t_1 = 0.7$\,ns and switched off for 
$\Delta t_2 = 0.266$\,ns (blue curve), compared with a continuous current of 
$I = 2.9 \times 10^{12}\,\mathrm{A/m^{2}}$ (red curve) that yields the same time-averaged driving. 
(b) Corresponding trajectories of the skyrmionium center, illustrating that pulsed excitation reproduces the net displacement of the dc-driven case while inducing short-time segments of reversed motion. 
(c,d) Analogous size evolution and trajectories for a smaller peak current 
$I_m = 3.0 \times 10^{12}\,\mathrm{A/m^{2}}$ and different ratios between the pulse-on and pulse-off intervals. 
By adjusting the duration of the zero-current interval, the effective time-averaged current is tuned, enabling access to distinct dynamical regimes without changing the peak current amplitude.
}
\end{figure}

To further analyze the deformed skyrmionium within the Mumax3 simulations and to
clarify the apparent deviation from the general result~(\ref{ideal}), we
numerically evaluated the components of the dissipative and gyrotropic tensors, together
with the resulting velocity components, for a driving current of
$I = 2.8 \times 10^{12}\,\mathrm{A/m^2}$, as discussed in the preceding section.
All micromagnetic simulations were performed using a 
uniform cell size of $1\,\mathrm{nm} \times 1\,\mathrm{nm} \times 1\,\mathrm{nm}$. 
The computed tensor elements are
\begin{equation}
D_{xx} = 67.37, D_{yy} = 86.45, D_{xy} = 0.127, G_{z} = -0.043.\nonumber
\end{equation}
These values yield the following velocity components and the Hall angle $\phi_{SHE}$, which characterizes the transverse deflection of the skyrmionium trajectory relative to the current flow:
\begin{equation}
v_x = 55.9~\mathrm{m/s}, v_y = -0.28~\mathrm{m/s}, \phi_{SHE}=-0.28^{\circ}.
\end{equation}

On the one hand, the small but finite skyrmionium Hall effect can be traced back
to discretization effects and numerical approximations involved in evaluating
the integrals defining the tensors $\mathbf{D}$ and $\mathbf{G}$.
Because the integrand in Eq.~(\ref{charge}) corresponds to an oriented
parallelogram on the unit sphere, a slight mismatch arises between the sphere
coverage associated with the inner skyrmion core and the corresponding
``uncovering'' contribution from the surrounding ring-shaped region.
As a result, the cancellation between opposite topological contributions is
not exact and a small residual gyrotropic component $G_z$ survives.
Even such a minute gyrotropic response is sufficient to induce a nonvanishing
transverse velocity $v_y$, thereby giving rise to a weak skyrmionium Hall
effect. This provides a transparent explanation for the small but finite Hall
deflection observed in our numerical simulations.
Naturally, employing finer spatial discretization further suppresses this
effect, driving the Hall angle $\phi_{\mathrm{SHE}}\to 0$.
{\color{black}For instance, reducing the simulation cell size from
$0.8\,\mathrm{nm}$ to $0.6\,\mathrm{nm}$ decreases the Hall angle from
$0.222^\circ$ to $0.124^\circ$, demonstrating the expected convergence toward
the continuum limit.

The continuum description, however, represents an idealized situation.
In a real magnetic crystal the magnetization texture is carried by a finite
set of atomic spins, and the number of spins contained within a characteristic
magnetic length scale becomes an important parameter.
For example, in the prototypical chiral magnets MnSi and FeGe the helical
pitch is approximately $18\,\mathrm{nm}$ and $70\,\mathrm{nm}$, corresponding
to about 39 and 149 lattice spacings, respectively.
Consequently, realistic modeling of such systems requires a numerical
discretization that faithfully resolves the finite number of spins involved
in the magnetic texture. In this situation the cancellation between the
topological contributions of the inner and outer parts of a skyrmionium may
remain slightly incomplete, leading to a small but finite residual gyrotropic
response and thus to a weak Hall deflection.}

On the other hand, the finite skyrmionium Hall effect may become genuinely
relevant for skyrmioniums of intrinsically small size, such as those typically
stabilized in systems with interfacial (induced) Dzyaloshinskii--Moriya
interaction \cite{Romming2013}. {\color{black}The magnetic periodicities observed
in ultrathin films and bilayer systems with interfacially induced
Dzyaloshinskii--Moriya interaction typically range from approximately
$1$ to $15\,\mathrm{nm}$ \cite{Romming2013}. As the skyrmionium radius decreases,
the magnetization varies more rapidly in space and the angular difference
between neighbouring spins increases.

Such nanoscale textures are naturally described within discrete spin models,
where the magnetic energy and topology are expressed through scalar and vector
products of unit-length spins rather than through spatial derivatives of a
continuous magnetization field.
The discretized counterpart of Eq.~(\ref{functional}) takes the form
\begin{align}
&w =  J\,\sum_{<i,j>} (\mathbf{S}_i \cdot \mathbf{S}_j ) -\sum_{i} \mathbf{H} \cdot \mathbf{S}_i 
 \nonumber\\
&- D \, \sum_{i} \left(\mathbf{S}_i \times \mathbf{S}_{i+\hat{x}} \cdot \hat{x}
 + \mathbf{S}_i \times \mathbf{S}_{i+\hat{y}} \cdot \hat{y}
 + \mathbf{S}_i \times \mathbf{S}_{i+\hat{z}} \cdot \hat{z}\right),
\label{discrete}
\end{align}
where classical spins of unit length are defined on a three-dimensional cubic
lattice and $<i,j>$ denotes pairs of nearest-neighbor spins.
Within this formulation, the Dzyaloshinskii--Moriya constant $D$ determines the
period $p$ of modulated structures according to
$D/J = \tan(2\pi/p)$.
Conversely, in numerical calculations one may choose a desired modulation period
$p$ and define the corresponding value of the Dzyaloshinskii--Moriya constant.
For example, the value $D = 0.7265J$ corresponds to a one-dimensional spiral
state with a period of ten lattice spacings in zero magnetic field.
Thus, within this discretized description, the number of spins per period of the
characteristic spiral state becomes an important parameter. In this framework,
the cancellation between the inner and outer topological contributions of a
skyrmionium cannot be artificially tuned by adjusting the spatial resolution,
as would be possible in continuum simulations with arbitrarily refined meshes.
Instead, the balance is determined by the discrete lattice representation itself.

From the viewpoint of a discrete spin lattice, this breakdown can be interpreted
in terms of an enhanced local scalar spin chirality, defined via mixed products
of neighbouring spins,
\begin{equation}
    \chi_{ijk} = \mathbf{S}_i \cdot (\mathbf{S}_j \times \mathbf{S}_k),
\end{equation}
which provides a natural measure of local vorticity in discrete systems.
The total topological charge is obtained by summing the local vorticity over all
lattice sites and normalizing by $4\pi$,
\begin{equation}
    Q = \frac{1}{4\pi}\sum_i \chi_i .
\end{equation}
}
%
%
As a consequence, a finite residual gyrotropic term $G_z$ is expected to emerge,
leading to an enhanced transverse velocity $v_y$ and, consequently, to a larger
effective Hall angle $\phi_{\mathrm{SHE}}$.
This implies that for ultracompact skyrmioniums the discretization-induced
gyrotropic response may evolve from a negligible numerical artifact into a
physically relevant contribution that should be taken into account when
analyzing their current-driven dynamics.

{\color{black}We emphasize, however, that the goal of the present work is not to
systematically vary the lattice parameters or to investigate the dependence on
the number of spins per period. Rather, our intention is to highlight possible
physical and numerical origins of the small residual Hall response and to draw
the attention of the skyrmionics community to this effect.}

Additionally, the resulting Hall response is not fixed but can be continuously
tuned by an external magnetic field, which modifies the relative population
of spins in the inner and outer skyrmionic regions and thereby controls the
degree of cancellation between their opposite topological contributions.
In this context, it is worth noting that \texttt{mumax3} employs a lattice-based
formulation for computing the topological charge, which significantly enhances
the numerical accuracy compared to approaches relying on finite-difference
derivatives of the magnetization field \cite{kim2020quantifying}.
This implementation substantially reduces discretization-induced errors and
thus provides a more reliable estimate of the residual gyrotropic response in
skyrmionium textures, allowing one to directly account for the contribution to
$G_z$ arising from the imbalance in the number of spins between the inner and
outer skyrmionic regions.

As a representative example, we consider a skyrmionium approaching the critical
line $a$--$b$, for instance at $h=0$  and $k_u=0.78\, (K_u = 4.68\times10^{5}~\mathrm{J/m^3})$.
As illustrated in Fig.~\ref{fig01}(c), the inner skyrmion core becomes very small near this line 
and is close to collapse, signaling the transition from a skyrmionium to an
ordinary skyrmion.
Using the same numerical grid in the \textsc{MuMax3} simulations, we obtain the
following values of the dissipative and gyrotropic tensors, together with the
resulting Hall response for $I = 1.6 \times 10^{12}\,\mathrm{A/m^2}$:
\begin{align}
&D_{xx}=51.43, D_{yy}=51.92, D_{xy}=0.024, G_{z}=-0.24,\nonumber\\
&v_x=32.01~\mathrm{m/s}, v_y=-1.49~\mathrm{m/s}, \phi_{\mathrm{SHE}}=-2.7^{\circ}.\nonumber
\end{align}
These values clearly demonstrate behavior characteristic of conventional
skyrmions. In this regime, the topological contribution of the shrinking inner
skyrmion is almost negligible, and the entire gyrotropic response originates from
the outer ring, which effectively hosts the topology of a single skyrmion.
For $k_u=0.8$ ($K_u \approx 4.8\times10^{5}~\mathrm{J/m^3}$), the Hall effect
increases even further, yielding $\phi_{\mathrm{SHE}}=-5.3^{\circ}$ for the same
applied current magnitude.

For comparison, we consider an axially symmetric skyrmion at the same control
parameters, $k_u=0.78, h=0$.
In this case, the numerically extracted tensor elements are
\begin{equation}
D_{xx}=18.98, D_{yy}\approx D_{xx}, D_{xy}=0, G_z=-12.469.\nonumber
\end{equation}
These values yield the velocities
$v_x=63.16~\mathrm{m/s}$ and $v_y=-4.75~\mathrm{m/s}$, corresponding to a Hall
angle
$\phi_{\mathrm{SHE}}=-4.3^{\circ}$, in excellent agreement with previous
reports~\cite{Brearton}. 

For an axially symmetric skyrmionic texture with $D_{xx}=D_{yy}\equiv D$ and $D_{xy}=0$,
the Thiele equation reduces to simple analytic expressions for the velocity
components,
\begin{equation}
v_x
=\frac{G_z^{\,2}+\alpha\xi D^{\,2}}{G_z^{\,2}+\alpha^{2}D^{\,2}}\;u_x,
\qquad
v_y
=\frac{G_z D\,(\alpha-\xi)}{G_z^{\,2}+\alpha^{2}D^{\,2}}\;u_x .
\end{equation}
Accordingly, the skyrmion Hall angle is given by
\begin{equation}
\tan\phi_{\mathrm{SHE}}
=\frac{v_y}{v_x}
=\frac{G_z D\,(\alpha-\xi)}
     {G_z^{\,2}+\alpha\xi D^{\,2}} .
\end{equation}

\begin{figure*}
  \centering
  \includegraphics[width=0.8\linewidth]{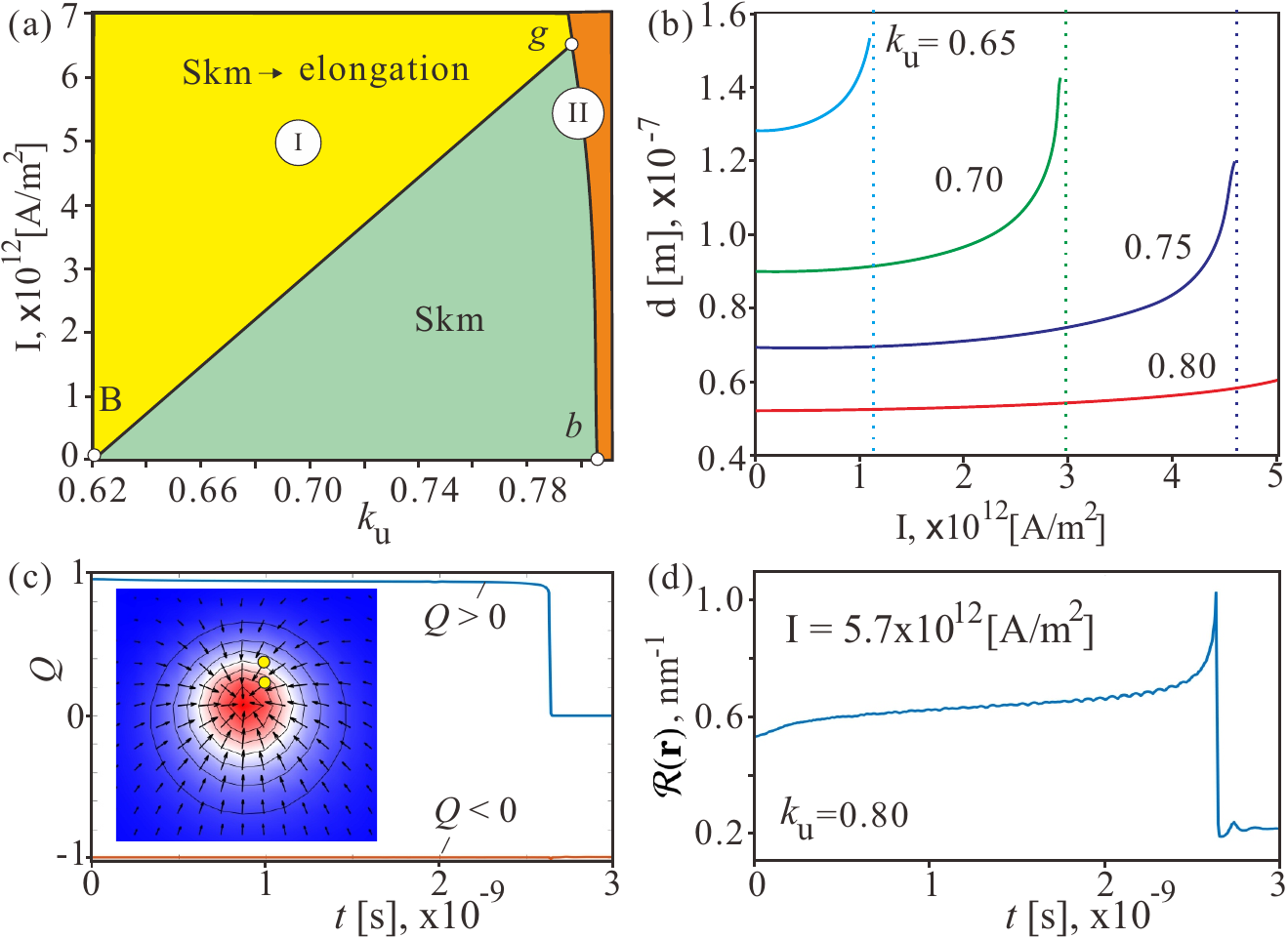}
  \caption{\label{fig04}
  (a) $k_u$–$I$ diagram showing the current-induced instability regimes of an isolated skyrmionium.
  In the yellow-shaded region~I, the applied current induces an unbounded elongation of the skyrmionium,
  whereas in the green-shaded region the skyrmionium remains stable, retaining an elongated yet finite profile.
  The orange-shaded region corresponds to the collapse of a skyrmionium into a skyrmion.
  (b) Skyrmionium size as a function of the driving current density for several values of the uniaxial anisotropy $k_u$.
  The critical current densities associated with the onset of unbounded elongation are indicated by dotted lines.
  For $k_u=0.80$, the terminal point of the red curve marks the transformation of the skyrmionium into a skyrmion.
  (c) Time evolution of the positive and negative contributions to the total topological charge during the collapse process.
  The inset shows a zoomed-in spin configuration of the internal skyrmion, with yellow markers indicating point defects.
  (d) Time evolution of the maximum value of the local rotational intensity $\mathcal{R}(\mathbf{r})$ [Eq.~(\ref{RR})]
  during the transformation.
  }
\end{figure*}

\textcolor{black}{To summarize, the skyrmionium Hall effect has a dual origin. 
In micromagnetic simulations, it can arise from discretization-induced imbalance of topological contributions and may be reduced by refining the computational grid, particularly in regions of rapidly varying magnetization. 
At the same time, in discrete spin systems with a finite number of spins per characteristic length scale, a small residual Hall response may persist and acquire physical relevance. In this case, it becomes sensitive to material parameters and can be tuned, for example, by varying the magnetic field, which modifies the internal structure of the skyrmionium. 
We therefore caution that such a residual Hall response is generically present in numerical simulations and may obscure other effects if not properly accounted for. At the same time, this sensitivity suggests that the skyrmionium Hall effect can be deliberately exploited and tuned, offering an additional degree of control in skyrmionium-based applications.}

\subsection{Response of isolated skyrmioniums to AC current driving}

As implied by Fig.~\ref{fig02}, the chiral spin texture of a skyrmionium remains
stable—although progressively elongated—and well preserved only up to a critical current
density of $I_{cr}\approx2.9\times10^{12}\,\mathrm{A/m^{2}}$, at which the maximal
attainable velocity reaches approximately $58\,\mathrm{m/s}$.
For larger driving currents, however, the internal forces acting on the two
constituent parts of the skyrmionium become excessively strong and eventually
tear the texture apart, leading to its self-destruction.
For $I \approx 3.0 \times 10^{12}\,\mathrm{A/m^{2}}$, the inset of
Fig.~\ref{fig02}(a) shows a continuously increasing skyrmionium size that no
longer saturates, in contrast to the behavior observed at lower current
densities.
This behavior is generic and signals proximity to the stability region of the
spiral state, i.e., to the line $c$--$B$ in the phase diagram
[Fig.~\ref{fig01}(e)].

To prevent such an instability event and to introduce an additional degree of control over Skm dynamics, we consider pulsed currents characterized by time intervals $\Delta t_1$, during which the current takes a constant value $I_m$, and $\Delta t_2$, during which the current is switched off [inset in Fig. \ref{fig08} (a)] \cite{zhang2022,ishida2020}. In this setup, the peak current densities $I_m$ can exceed the critical current $I_{\mathrm{cr}}$ while avoiding a continuous overdriving that would otherwise lead to Skm elongation instability.

As a result, an accidental increase of the driving current does not necessarily trigger the instability, and different dynamical characteristics in this regime are governed by the time-averaged current, defined as
\begin{equation}
\langle I \rangle = \frac{1}{\Delta t_1 + \Delta t_2}
\int_0^{\Delta t_1 + \Delta t_2} I(t)\, dt
= I_m \frac{\Delta t_1}{\Delta t_1 + \Delta t_2}.
\end{equation}
Thus, the effect of the pulsed-current protocol can be directly compared to that of a continuous driving current with magnitude $\langle I \rangle$.

Figure~\ref{fig08}(a) shows the time evolution of the Skm size for a pulsed current with
$I_m = 4.0 \times 10^{12}\,\mathrm{A/m^{2}}$, $\Delta t_1 = 0.7\,\mathrm{ns}$, and
$\Delta t_2 = 0.266\,\mathrm{ns}$ (blue line). These parameters correspond to a continuous driving current of
$I = 2.9 \times 10^{12}\,\mathrm{A/m^{2}}$ (red line) as defined by the time-averaged current.
Despite the overall agreement between the pulsed and continuous driving protocols, the Skm displacement
under pulsed currents exhibits distinct trajectory segments in which the Skm transiently moves in the
opposite direction [blue line in Fig. \ref{fig08}(b)].

Figures~\ref{fig08}(c,d) display the same characteristics as in panels~(a,b), but for different ratios of the pulse durations $\Delta t_1$ and $\Delta t_2$ and for a reduced peak current density $I_m = 3.0 \times 10^{12}\,\mathrm{A/m^{2}}$. An increase of the off-time interval $\Delta t_2$ is equivalent to a reduction of the time-averaged current $\langle I \rangle$. Consequently, for a fixed peak current density $I_m$, one can effectively access different dynamical regimes by varying $\Delta t_2$.

 \textcolor{black}{
To summarize, our goal here is to demonstrate that the instability mechanisms identified for DC currents can be mitigated under experimentally realistic pulsed driving conditions. While AC or pulsed current driving may in principle give rise to additional dynamical effects—such as nonlinear responses, frequency-dependent behavior, or resonance phenomena—a systematic exploration of these aspects lies beyond the scope of the present work. In this study, pulsed driving is primarily employed as a practical means of controlling the effective current delivered to the system while avoiding excessively large continuous currents. Within the parameter range considered here, the resulting dynamics is well described by the time-averaged current density and therefore closely resemble those obtained under DC driving.
}


\begin{figure*}
  \centering
  \includegraphics[width=0.8\linewidth]{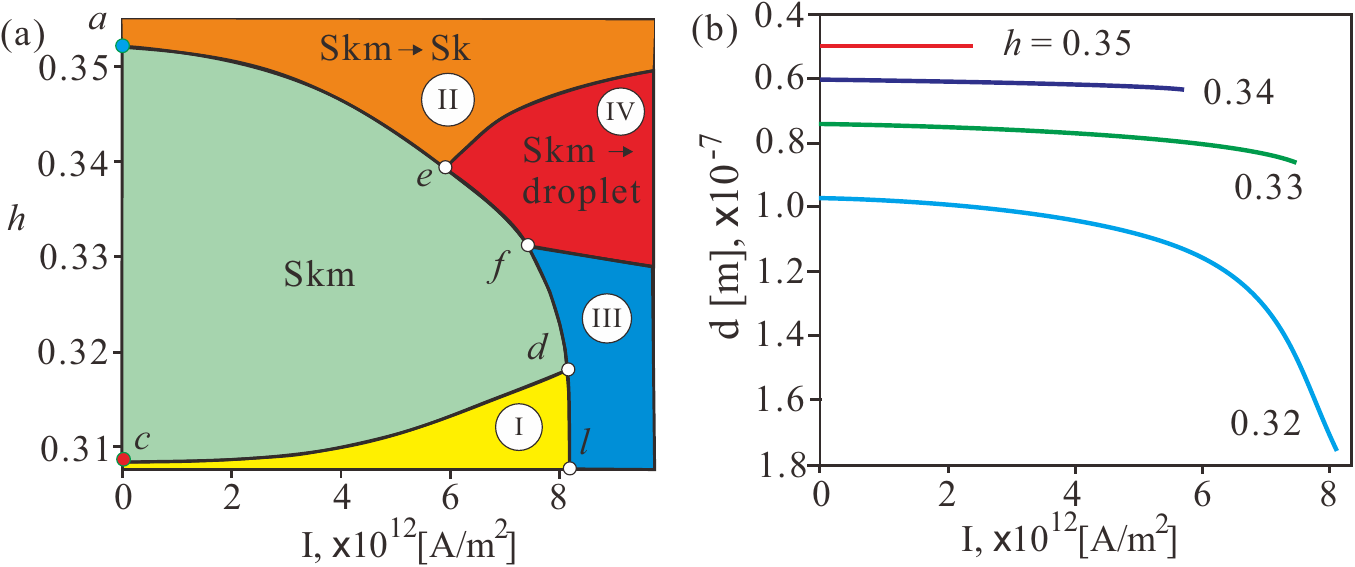}
  \caption{\label{fig03}
  (a) $h$–$I$ diagram for $k_u=0$, which, similarly to Fig.~\ref{fig04}(a), delineates
  the critical current densities and the corresponding instability regimes of
  isolated skyrmioniums.
  In regions~I and~II, a skyrmionium undergoes either current-induced elongation
  or collapses into an isolated skyrmion, respectively.
  In region~III, the skyrmionium expands to occupy the entire system through the
  formation and elongation of meron-tipped stripe textures emerging from the
  skyrmionium boundary.
  In region~IV, a skyrmionium transforms into a topologically trivial magnetic
  droplet, which can be viewed as a composite object consisting of half a
  skyrmion and half an antiskyrmion.
  (b) Skyrmionium size as a function of the driving current density for several
  values of the applied magnetic field.
  The termination points of the curves lie along the line $a$–$e$–$f$–$d$ in (a).
  }
\end{figure*}

\section{Current-induced transformations and instabilities of skyrmioniums}

In the present section, we examine in detail additional instability mechanisms encountered by isolated skyrmioniums under current-driven dynamics.
We stress that the current-induced degradation processes discussed below should not be viewed merely as failure modes; instead, they offer valuable insight into the underlying topological energy landscape that governs chiral spin textures. {\color{black} Thus, our goal is not to determine precise numerical values of the critical currents for specific materials, but rather to use skyrmioniums as probes of the underlying magnetic energy landscape. 
By studying how the textures transform under applied currents, one can identify the preferred pathways through which the system evolves toward energetically more stable magnetic states at fixed external parameters. In this sense, the applied current acts as a tool for inducing controlled out-of-equilibrium transformations that reveal the connectivity between different magnetic states in the energy landscape.}

\begin{figure}
  \centering
  \includegraphics[width=0.99\linewidth]{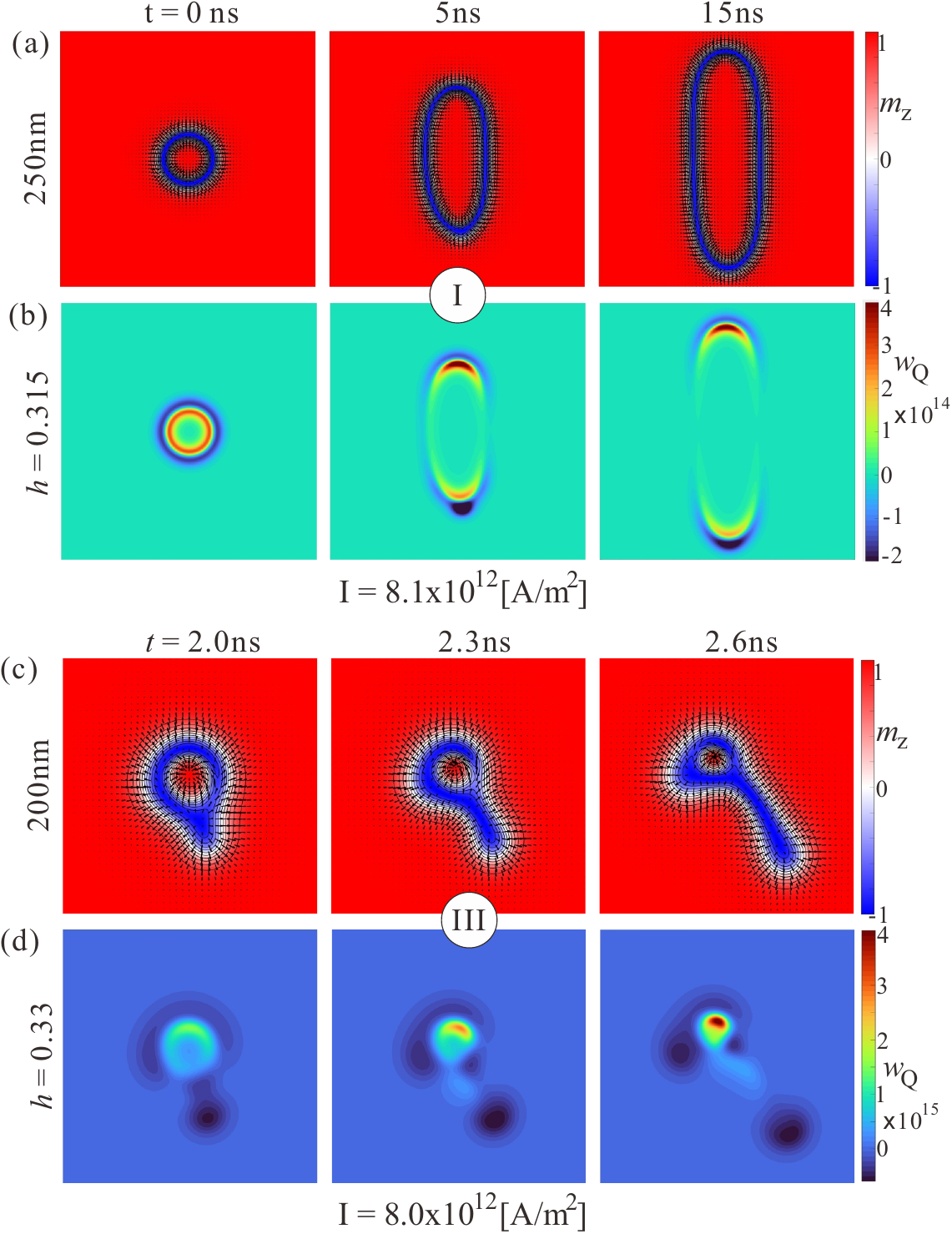}
  \caption{\label{fig05}
  (a,b) Sequence of snapshots illustrating the elliptical (elongation) instability
  of an isolated skyrmionium occurring along the line $c$–$d$ of the $h$–$I$
  diagram shown in Fig.~\ref{fig03}(a) for $h=0.315$ and $k_u=0$.
  The upper row shows color maps of the $m_z$ component of the magnetization,
  while the lower row displays the corresponding topological charge density
  $w_Q(x,y)$.
  (c,d) Sequence of snapshots illustrating the instability mechanism mediated by
  meron-tipped stripe formation occurring within region~III of
  Fig.~\ref{fig03}(a).
  }
\end{figure}

\subsection{Instabilities of skyrmioniums in the ferromagnetic regime}

\textcolor{black}{In this section, we investigate current-driven instabilities of isolated skyrmioniums in the ferromagnetic region of the phase diagram at zero magnetic field while varying the uniaxial anisotropy $k_u$. 
}

The $k_u-I$ diagram in Fig.~\ref{fig04}(a) shows a nearly linear increase of the critical
current associated with the elongation instability, starting from the point
$k_u(B)$ (point $B$ at the phase diagram in Fig. \ref{fig01}(e)) and terminating at $k_u(g)\approx 0.795$.
In the yellow-shaded region labeled~I, the applied current induces an unbounded
elongation of the skyrmionium, whereas in the green-shaded region the
skyrmionium remains stable, retaining an elongated yet finite profile.
Figure~\ref{fig04}(b) quantifies the skyrmionium size as a function of the driving current density for several values of the anisotropy, in close analogy with Fig.~\ref{fig02}(a).

Remarkably, with increasing uniaxial anisotropy $k_u$, the elongation instability
of skyrmioniums is replaced by a collapse into an isolated skyrmion, signaling
close proximity to the $a$--$b$ line in the phase diagram.

As shown in Fig.~\ref{fig04}(b), the overall size of the skyrmionium remains
nearly unchanged until the internal skyrmion with positive polarity collapses
(red curve). The orange-shaded collapse region is labeled~II in the diagram shown in
Fig.~\ref{fig04}(a).

The collapse of the skyrmionium is essentially instantaneous, as evidenced by
Fig.~\ref{fig04}(c), which displays the time evolution of the positive and
negative contributions to the topological charge within the skyrmionium (see
also Supplementary Video~\#1).

To identify regions where the magnetization exhibits extremely rapid spatial
variations, we introduce a scalar measure of the local rotational intensity,
defined as
\begin{equation}
\mathcal{R}(\mathbf{r}) =
\sqrt{
\left|\frac{\partial \mathbf{m}}{\partial x}\right|^{2}
+
\left|\frac{\partial \mathbf{m}}{\partial y}\right|^{2}
}
=
\sqrt{
\sum_{\mu=x,y,z}
\left[
\left(\frac{\partial m_\mu}{\partial x}\right)^2
+
\left(\frac{\partial m_\mu}{\partial y}\right)^2
\right]
},
\label{RR}
\end{equation}
where $\mathbf{m}=(m_x,m_y,m_z)$ is the unit magnetization field.

Small values of $\mathcal{R}$ correspond to smoothly varying spin textures,
such as the uniform ferromagnetic state or the far-field region of solitons.
Moderate values are characteristic of rapid but continuous rotations,
for example within domain walls or along the boundary of a skyrmionium.
In contrast, very large values of $\mathcal{R}$ indicate regions of extreme
noncollinearity, such as defect-like points or collapse events.

In the continuum limit, $\mathcal{R}(\mathbf r)$ is unbounded and diverges at
true singularities of the magnetization field.
In numerical simulations, however, its maximal value is constrained by the mesh
spacing: for a square grid with cell size $\Delta_{x}=\Delta_{y}=a$,
\begin{equation}
\mathcal{R}_{\max} \sim \frac{2\sqrt{2}}{a},
\end{equation}
corresponding to nearly antiparallel spins on neighboring cells.
For the present simulations with $a=1\,\mathrm{nm}$,
this yields $\mathcal{R}_{\max}\sim2.8\,\mathrm{nm^{-1}}$.

Local defect points are therefore identified as positions satisfying
\begin{equation}
\mathcal{R}(\mathbf{r}) > \mathcal{R}_{\mathrm{c}},
\end{equation}
where $\mathcal{R}_{\mathrm{c}}$ is a prescribed threshold.
Such regions represent strong local noncollinearity and can be interpreted as
Bloch-point–like or discontinuity-like defects arising from the discrete
numerical representation of the spin texture.

In the inset of Fig.~\ref{fig04}(c), points with
$\mathcal{R}(\mathbf{r}) > 1$ are highlighted by yellow circles.
Figure~\ref{fig04}(d) shows the time evolution of the maximal value of
$\mathcal{R}(\mathbf{r})$ during the transformation process.

Interestingly, a closer inspection of the $k_u-I$ diagram in Fig.~\ref{fig04}(a) reveals
a crossover regime in which both instability mechanisms coexist. To the left of
point~$g$, a skyrmionium undergoes current-induced elongation at lower current
densities, while at higher currents it collapses into an isolated skyrmion.
Overall, since the entire interval $B$--$b$ corresponds to the ferromagnetic
region of the phase diagram, all extended magnetization textures must eventually
shrink back to either a skyrmionium or a skyrmion once the driving current is
switched off.

\begin{figure*} \centering \includegraphics[width=0.8\linewidth]{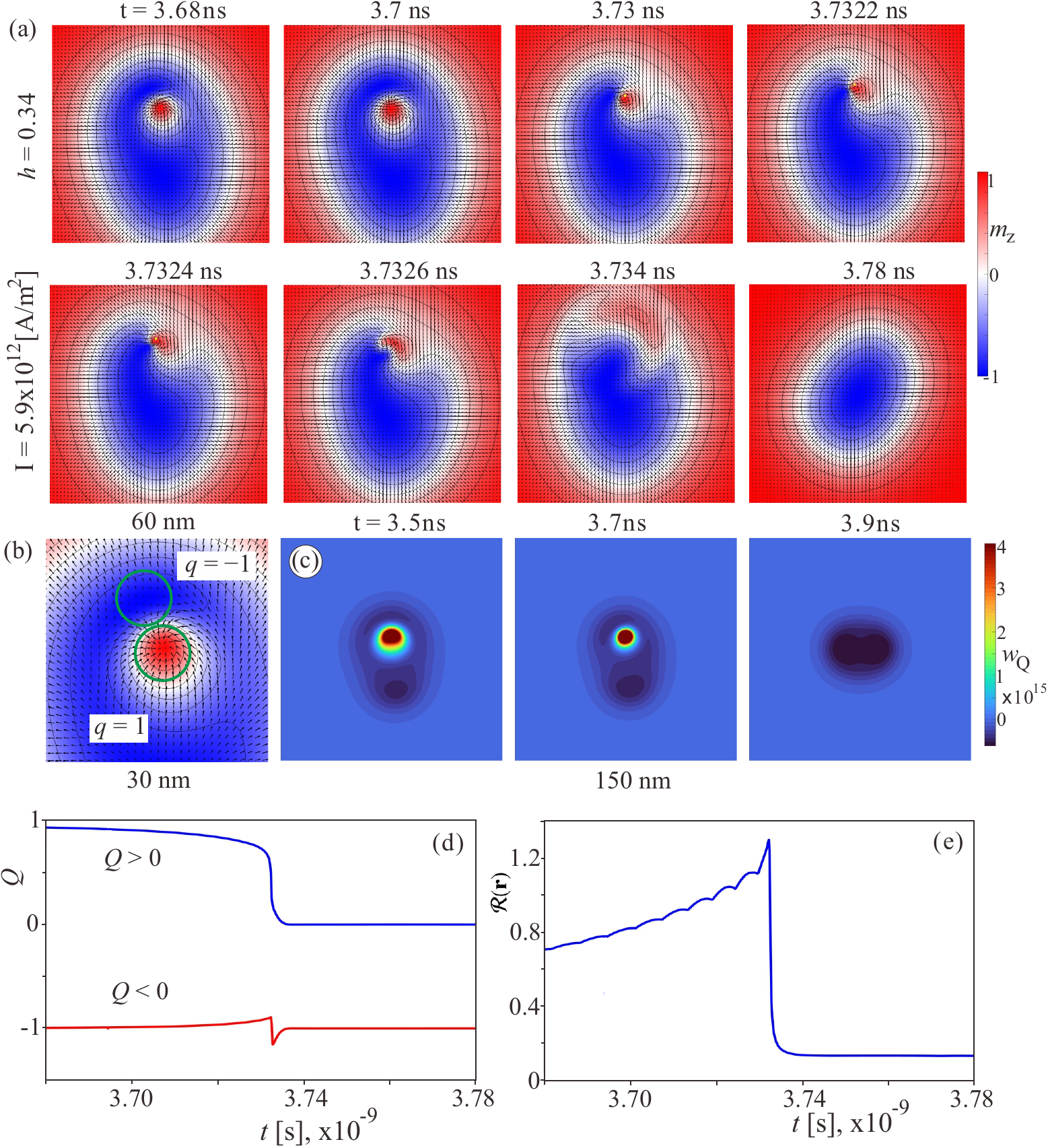} \caption{\label{fig06} Collapse of a skyrmionium into a skyrmion under a dc current. (a) Time sequence of snapshots showing the evolution of the $m_z$ component (color scale) with in-plane magnetization projections indicated by black arrows for $h=0.34$ and $I=5.9\times10^{12}\,\mathrm{A/m^2}$. The inner skyrmion becomes asymmetrically exposed to the ferromagnetic background, leading to its effective expulsion from the skyrmionium interior. (b) Intermediate stage of the transformation, where two localized (meron) magnetization textures with opposite winding numbers (marked by green circles) emerge. (c) Corresponding evolution of the topological charge density, demonstrating that both textures carry a positive contribution to the total topological charge. (d) Time dependence of the positive and negative parts of the total topological charge, illustrating their mutual annihilation during the collapse process. (e) Temporal evolution of the maximal local rotational intensity $\mathcal{R}$ [Eq.~(\ref{RR})], indicating the involvement of point-defect–like singularities during the skyrmionium-to-skyrmion transformation. } \end{figure*} \begin{figure*} \centering \includegraphics[width=0.7\linewidth]{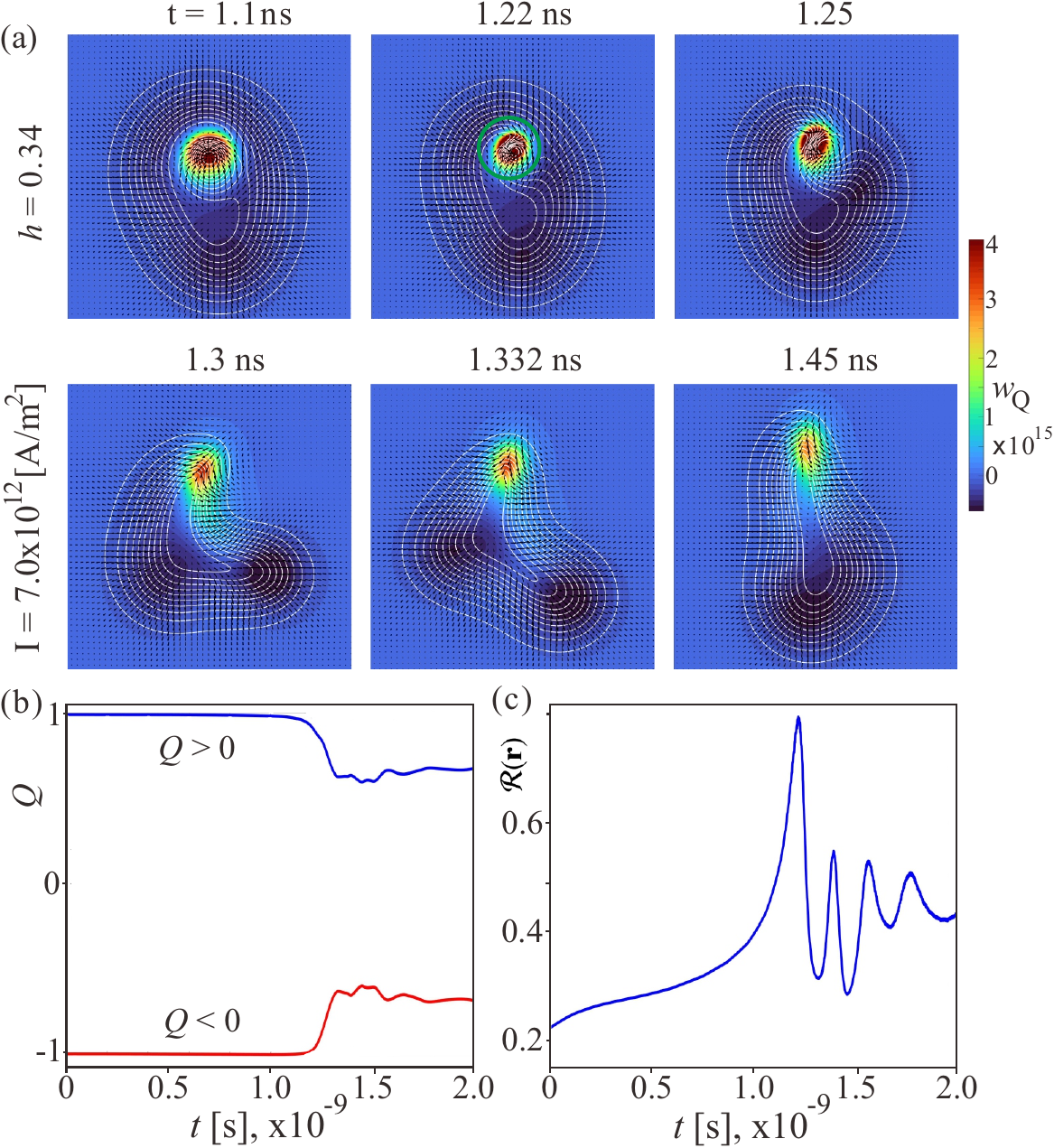} \caption{\label{fig07} Current-induced transformation of a skyrmionium into a topologically trivial droplet (region~IV in Fig.~\ref{fig03}(a)). (a) Sequence of snapshots showing the topological charge density (color scale), with in-plane magnetization directions indicated by arrows and white contours marking constant $m_z$ levels. The images illustrate how the inner skyrmion component gradually connects to the ferromagnetic background and is released, in contrast to the collapse mechanism depicted in Fig.~\ref{fig06}. (b) Time evolution of the positive and negative contributions to the total topological charge, demonstrating the progressive loss of approximately one half of the positive charge and the formation of a droplet composed of a half-skyrmion and a half-antiskyrmion, while the total topological charge remains zero throughout the process. (c) Time dependence of the maximal local rotational intensity $\mathcal{R}$ [Eq.~(\ref{RR})], which stays below $\sim0.8\,\mathrm{nm^{-1}}$, indicating that the transformation proceeds without the creation of singular point defects and is governed by smoothly distributed regions of rapid magnetization rotation. } \end{figure*}

{\color{black}Thus, the instability mechanisms considered here illustrate how the current-driven dynamics reflects the structure of the equilibrium phase diagram: depending on the position within the diagram, the system follows different instability pathways toward energetically favored states.

For relatively small values of the anisotropy parameter $k_u$, the skyrmionium resembles a flexible circular domain-wall ring. In this regime, the current-induced spin torque primarily causes an elongation of the structure, eventually transforming the skyrmionium into an extended stripe-like configuration.
For larger values of $k_u$, the inner skyrmion core within the skyrmionium becomes progressively smaller and approaches the collapse threshold. In this regime, the dominant current-induced instability is the collapse of the inner skyrmion, resulting in the transformation of the skyrmionium into a single skyrmion.
The two instability lines observed in the phase diagram therefore correspond to these distinct mechanisms and merge at an intermediate value of $k_u$, marking the crossover between the domain-wall–dominated and core-collapse regimes.}

\subsection{Skm transformations within the region of stable SkL}

Within the region $c$--$a$--$D$--$B$ of the phase diagram, the metastability region
of skyrmioniums overlaps with the stability region of the hexagonal skyrmion
lattice. In this section, we investigate the consequences of this overlap for
the current-induced instabilities of isolated skyrmioniums.

Figure~\ref{fig03}(a) presents the $h$–$I$ diagram for $k_u=0$, which, similarly
to Fig.~\ref{fig04}(a), delineates the critical currents and the corresponding
instability regimes, but exhibits a richer structure. Depending on the location
within the phase diagram, a skyrmionium can undergo distinct current-driven
transformation pathways, which can be systematically classified.

As before, one identifies regions~I and~II, in which a
skyrmionium undergoes either current-induced elongation or collapses into an
isolated skyrmion, respectively. {\color{black} As the magnetic field increases, the skyrmionium shrinks and the central skyrmion core approaches the collapse threshold. Consequently, smaller current densities are sufficient to trigger collapse-type instabilities at larger magnetic fields, as reflected by the boundary along the line $a$--$e$. In contrast, at decreasing magnetic fields the skyrmionium forms a relatively large and flexible circular domain-wall ring. In this regime, lower current densities are required to overcome the elastic stiffness of the domain-wall boundary and induce strong deformations such as elongation along the line $c$--$d$. Overall, even in the absence of an applied current, the skyrmionium becomes unstable at the boundaries of its metastability region: it collapses at point $a$ or elongates at point $c$.}

In Fig.~\ref{fig05}(a,b), we quantify the temporal elongation instability of an
isolated skyrmionium at $k_u=0$ and $h=0.315$ ($\mu_0 H=0.33~\mathrm{T}$), driven by
a current density $I=8.1\times10^{12}\,\mathrm{A/m^{2}}$.
In the first row, we present color maps of the $m_z$ component of the
magnetization.
In the second row, the corresponding topological charge density $w_Q$ is shown,
revealing an imbalance of positive (negative) topological charge accumulating
in the upper (lower) part of the texture. In the $h-I$ diagram (Fig.~\ref{fig03}(a)), the line $c$--$d$ corresponds to this elongation process.

Interestingly, at larger current densities and within essentially the same range
of applied magnetic fields (namely, along the line $f$--$d$--$l$), the
skyrmionium no longer undergoes elongation but instead transforms into a
disordered state composed of stripe-like textures spanning the entire system
(blue-shaded region~III in Fig.~\ref{fig03}(a)).
During this process [Fig.~\ref{fig05}(c,d)], a meron-tipped stripe protrudes from
the skyrmionium boundary and elongates \cite{kuchkin2023}, while a second stripe
simultaneously develops on the opposite side, ultimately leading to the collapse
of the central skyrmion (see Supplementary Video~\#2).
In Ref.~\cite{kuchkin2023}, such solitonic objects were termed \emph{tailed
skyrmions} and were shown to emerge within a narrow field interval close to the
transition between the spiral and field-polarized states.
From this perspective, the above scenario can be interpreted as a
current-driven excitation of a skyrmionium into a stick-like (tailed) skyrmion.

One can take a step further toward understanding the physical origin of this
transformation.
Because the corresponding parameter range in the equilibrium phase diagram
[Fig.~\ref{fig01}(e)] lies inside the stability region of the skyrmion lattice, as
discussed above, the emergent stripe-like textures—together with the
infinitely elongated skyrmionium loops shown in Fig.~\ref{fig05}(a,b)—can serve
as effective nucleation media for skyrmions.
Once the driving current is switched off, the resulting magnetic configuration
does not relax back into an isolated soliton.
Instead, local ruptures of the stripe pattern may generate individual skyrmions,
which subsequently assemble into a lattice with the equilibrium period. {\color{black} In other words, the system tends to fill space with energetically favored topological kinks, which can occur via two related pathways (mechanisms I and III): either through the global elongation of the skyrmionium or through the local formation of a stripe-like protrusion that develops from the skyrmionium boundary.}
A closely related nucleation mechanism was previously reported in
Ref.~\cite{Mukai}.

Near the line $a$--$b$ in the phase diagram [Fig.~\ref{fig01}(e)], where a
skyrmionium approaches collapse into an isolated skyrmion, one can readily
anticipate that an applied electric current may facilitate this
transformation. Such a current-assisted collapse occurs along the line $a$--$e$
in the $h$–$I$ diagram shown in Fig.~\ref{fig03}(a), corresponding to the
orange-shaded region.
Interestingly, at still larger current densities the skyrmionium does not
follow this collapse pathway but instead undergoes a qualitatively different
instability: it transforms into a topologically trivial magnetic droplet \cite{sisodia2021,zhou2015,rozsa2017,leonov2021}, which can be viewed as a composite object consisting of
half a skyrmion and half an antiskyrmion, yielding a total topological charge
$Q=0$. This regime is indicated by the red-shaded region labeled~IV in Fig.~\ref{fig03}(a). {\color{black}Thus, mechanism IV represents an intermediate scenario. Here, the skyrmionium approaches collapse into a skyrmion, as in mechanism II, but instead follows a topologically allowed pathway toward a topologically trivial droplet state. When the current is switched off, the droplet subsequently collapses into the homogeneous state. In this way, instability mechanism IV reveals a previously unexplored topological pathway connecting the skyrmionium to the homogeneous state, consistent with the fact that both configurations carry trivial total topological charge. The applied current facilitates this transformation by enabling the system to overcome the potential barrier separating the two states.}

{\color{black}
To summarize, the first two instability mechanisms discussed above—elongation of the skyrmionium and collapse into a single skyrmion—are qualitatively similar to those identified previously in the zero-field case with varying anisotropy. They reflect intrinsic deformation modes of the skyrmionium texture itself. In contrast, the additional intermediate mechanisms observed in the present $h$–$I$ diagram arise from the fact that the considered parameter range lies inside the equilibrium stability region of the skyrmion lattice. In this situation, the current-driven dynamics does not simply destabilize the isolated skyrmionium but also promotes the nucleation of the energetically favored periodic phase. As a result, the skyrmionium can emit stripe-like protrusions that act as precursors of extended modulated textures, or transform into a droplet-like state that subsequently decays toward configurations compatible with the skyrmion-lattice environment. These processes therefore represent current-assisted pathways through which the system evolves from an isolated metastable soliton toward the spatially extended structures preferred by the equilibrium phase diagram.}

In the following, we analyze the two collapse processes separately, since their microscopic evolution differs qualitatively from the collapse scenario illustrated in Fig.~\ref{fig04}(c).

\subsubsection{Collapse of a skyrmionium into a skyrmion}

A representative example of the transformation of a skyrmionium into a skyrmion is shown in Fig.~\ref{fig06}(a) for $h=0.34$ and $I=5.9\times10^{12},\mathrm{A/m^{2}}$. The process is illustrated by a sequence of snapshots displaying color maps of the $m_z$ component of the magnetization, with black arrows indicating the in-plane ($xy$) projections (see also Supplementary Video~\#3).
Initially, the central skyrmion becomes progressively exposed to the surrounding ferromagnetic background from the right-hand side. In this region, the magnetization rotation from the skyrmion core to the background does not reach $m_z=-1$, in contrast to the left-hand side. As time evolves, the maximal negative value of $m_z$ steadily increases, giving the impression that the central skyrmion is being expelled from the interior of the skyrmionium.
Subsequently, two distinct magnetization patterns with opposite winding numbers
emerge (highlighted by green circles in Fig.~\ref{fig06}(b)).
Despite their opposite winding, both textures carry a positive topological
charge, as shown in Fig.~\ref{fig06}(c).
The resulting composite object can therefore be identified as a bimeron, with a
total unit topological charge distributed between two merons.

At a later stage (snapshot at $t=3.734\mathrm{ns}$), these two components mutually annihilate, resulting in the complete removal of the positive topological charge initially contained within the skyrmionium profile. Figure~\ref{fig06}(d) presents the time evolution of the positive and negative contributions to the total topological charge. Figure~\ref{fig06}(e) shows the temporal behavior of the scalar measure of the local rotational intensity $\mathcal{R}$ (Eq. (\ref{RR})), providing clear evidence for the involvement of point-defect–like singularities in the collapse process.

Thus, while both collapse processes shown in Figs.~\ref{fig04}(e) and \ref{fig06} ultimately lead to the disappearance of the central skyrmion and formation of defects, they proceed via distinct microscopic mechanisms and follow qualitatively different dynamical pathways.

\subsubsection{Transformation of a skyrmionium into a droplet}

The skyrmionium instability with respect to the droplet \cite{sisodia2021,zhou2015,rozsa2017} proceeds through the same initial stages as those depicted and described in Fig.~\ref{fig06}, at the same magnetic field but at a slightly larger current density. The red-shaded region~IV in Fig.~\ref{fig03}(a) corresponds to this transformation.

Instead of the collapse of the previously described pair with opposite winding numbers (encircled by two green circles), the counterpart corresponding to the skyrmion core with $q=+1$ gradually connects to the ferromagnetic background and is effectively ``released'' (Fig. \ref{fig07}(a)), thereby losing almost half of the total positive topological charge, as shown by the temporal evolution in Fig.~\ref{fig07}(b). The remaining half of the positive topological charge associated with the $q=-1$ counterpart persists and forms the upper tip of the emerging droplet. Eventually, the lower part of the droplet accommodates the shape of a half-skyrmion, which carries a partial negative topological charge, as also indicated in Fig.~\ref{fig07}(b). Throughout this process, the total topological charge remains zero, as in the original skyrmionium configuration (see also Supplementary Video~\#4).

Remarkably, the value of $\mathcal{R}$ [Eq.~(\ref{RR}), Fig.~\ref{fig07}(c)] does not exceed $0.8\mathrm{nm}^{-1}$, indicating that the transformation proceeds predominantly via regions of rapid rotation rather than through the formation of singular defects. This behavior suggests that the numerical accuracy of this process could be further improved in finite-difference schemes employing smaller cell sizes.

With a further increase in the driving current, the droplet soliton undergoes pronounced elongation and eventually expands to fill the entire system with stripe-like textures, thereby providing an additional nucleation pathway for the formation of a skyrmion lattice.

Such a gradual unwinding of a skyrmionium into a droplet \cite{leonov2021} offers an
alternative perspective on the internal structure of skyrmioniums. Rather than
viewing a skyrmionium as two nested skyrmions, it may be interpreted as a swirled
droplet state which, via the annihilation of its tips carrying opposite winding
numbers (and topological charges), can transform back into a skyrmionium.
Although both the skyrmionium and the homogeneous state possess zero total
topological charge, a smooth adiabatic transformation between them is not
energetically trivial.
In particular, the droplet configuration necessarily involves the nucleation of
two terminal defects with opposite winding numbers, which must first be created
and subsequently separated before annihilation can occur.
The formation and spatial separation of these charged endpoints entails a finite
exchange and DMI energy cost, thereby giving rise to an effective potential
barrier that stabilizes the skyrmionium as a metastable state even in the absence
of topological protection \cite{leonov2026metamatter}.

{\color{black} To summarize, our motivation in studying skyrmionium instabilities is to use skyrmioniums as probes of the underlying magnetic energy landscape. By analyzing how they transform under applied currents, one can identify the preferred pathways through which the system evolves toward energetically more stable magnetic states at the same external parameters. Our emphasis is therefore on revealing the interconnections between different magnetic states and the associated instability mechanisms, rather than on the precise numerical values of the critical currents at which individual transformations occur. In this sense, we use the applied current as a tool to induce controlled out-of-equilibrium transformations that help elucidate the structure of the underlying energy landscape.
Remarkably, the evolution of skyrmioniums under applied current also provides information about the system’s position within the magnetic phase diagram, thereby allowing one to infer, at least qualitatively, the corresponding regime of material parameters, such as strong or weak anisotropy.

Moreover, the qualitative form of different instability pathways discussed in the preceding sections can be understood from the competition between the fundamental energy contributions that determine the equilibrium structure of the skyrmionium, namely the exchange, DMI, Zeeman, and anisotropy energies. Their interplay controls the characteristic size and internal profile of the skyrmionium. In particular, for relatively small magnetic fields the skyrmionium forms a large circular domain-wall ring whose stability is primarily governed by the balance between exchange and DMI energies that determine the domain-wall tension. In this regime, current-induced spin torques mainly excite deformations of the domain-wall boundary, leading to elongation or stripe-like protrusions. By contrast, at larger magnetic fields the Zeeman energy favors a reduction of the core, causing the inner skyrmion within the skyrmionium to shrink and approach the collapse threshold. In this case the dominant instability involves the collapse of the central skyrmion core, which may lead either to the formation of a single skyrmion or to a transient droplet-like configuration.
}

\begin{figure*}
  \centering
  \includegraphics[width=0.7\linewidth]{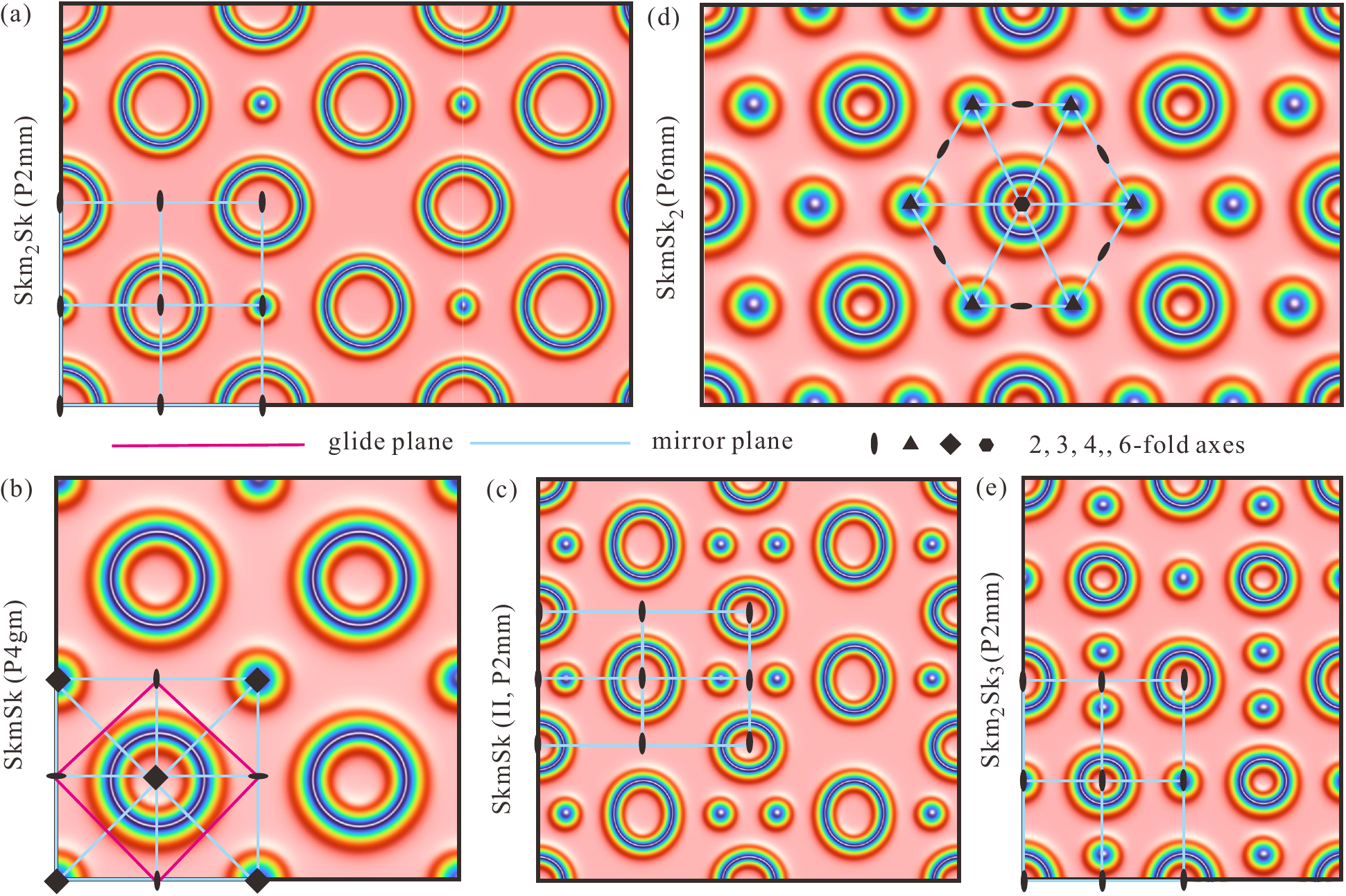}
  \caption{\label{fig09}
Examples of composite skyrmionium–skyrmion (Skm–Sk) lattices realizing distinct
topological compositions.
Panels (a)–(e) illustrate representative periodic arrangements with increasing
skyrmion content: (a) Skm$_2$Sk lattice (2:1), (b) SkmSk checkerboard structure
(1:1), (c) alternative SkmSk geometry in which two skyrmions are trapped inside
a plaquette formed by four surrounding skyrmioniums, (d) SkmSk$_2$ hexagonal
lattice (1:2), and (e) Skm$_2$Sk$_3$ lattice (2:3).
The color scale encodes the out-of-plane magnetization component $m_z$.
For each configuration, the corresponding plane-group symmetry is indicated,
emphasizing how the local coordination between skyrmioniums and skyrmions
governs the global lattice geometry.
  }
\end{figure*}

\section{Current-driven processes in skyrmionium-based meta-matter}

\subsection{The concept of Skm meta-matter}

Recent work~\cite{leonov2026metamatter} has revealed that a lattice composed solely of
skyrmioniums, which is expected to appear in proximity to the spiral phase
within the region $0$--$c$--$B$ of the phase diagram
[Fig.~\ref{fig01}(e)], is generically unstable.
Rather than forming a robust periodic state, the constituent skyrmioniums
develop an elliptical distortion that continuously drives the system toward
the energetically favorable spiral configuration.
This instability reflects the intrinsic tendency of extended skyrmionium
textures to lower their energy by elongation when competing with modulated
ground states.

These findings motivated the introduction of composite lattices in which
skyrmioniums coexist with conventional skyrmions, forming what was termed
\emph{skyrmionium-based meta-matter} in Ref.~\cite{leonov2026metamatter}.
In such mixed Skm–Sk crystals, objects with qualitatively different topology
($Q \approx 0$ for skyrmioniums and $Q=-1$ for skyrmions) are arranged within a
single periodic structure.
The resulting interplay between topologically trivial and nontrivial
constituents gives rise to effective interactions that can suppress the
elongation mode responsible for the instability of the pure skyrmionium
lattice, thereby stabilizing ordered states unattainable in homogeneous Skm
or Sk assemblies.

Figure~\ref{fig09} illustrates several representative realizations of these
Skm–Sk composite lattices, characterized by different topological
``stoichiometries,'' i.e., distinct ratios of skyrmioniums to skyrmions.
The associated spatial symmetries and plane groups are indicated, highlighting
the structural diversity achievable within this class of meta-matter.

Beyond their static stability, mixed Skm–Sk lattices were shown in
Ref.~\cite{leonov2026metamatter} to host a complex spectrum of collective excitations.
These modes arise from the hybridization of internal oscillations of the
individual constituents with long-range interactions mediated by the lattice,
leading to a hierarchical magnonic response.
As a result, skyrmionium-based meta-matter emerges as a promising platform for
magnonic functionality, where the coupling between distinct topological
quasiparticles can be tailored to engineer tunable and multifunctional
spin-wave dynamics. {\color{black} In addition, the reduced transverse motion of skyrmioniums under current makes them attractive candidates for controlled transport in racetrack-type geometries. }

\begin{figure}
  \centering
  \includegraphics[width=0.99\linewidth]{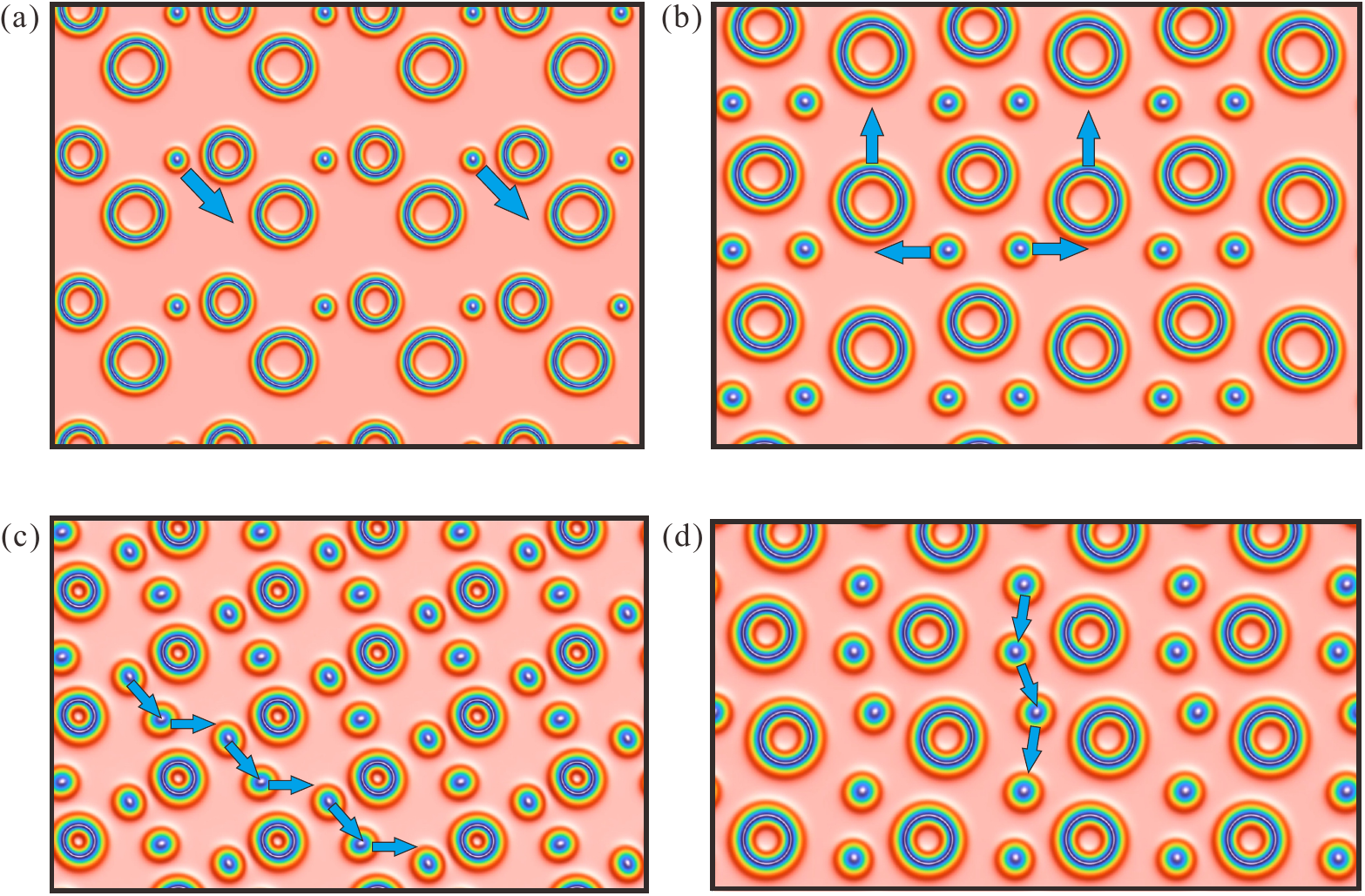}
  \caption{\label{fig10}
Current-driven dynamics of skyrmionium metamatter. Panels (a)--(d) show color maps of the $m_z$ component of the magnetization corresponding to Supplementary Videos~5, 6, 8, and 9, respectively. Blue arrows indicate the directions of motion of skyrmioniums and skyrmions (see text for details).
  }
\end{figure}

{\color{black} Skyrmionium metamatter could, in principle, be realized in the same classes of
chiral magnetic materials in which skyrmion lattices are known to stabilize
within a comparable parameter window. From a practical perspective, one
possible route would be to start from a skyrmion lattice and locally transform
selected skyrmions into skyrmioniums, thereby creating a prescribed spatial
pattern. Similar local writing procedures have already been demonstrated for
skyrmions in ultrathin magnetic films, for example in PdFe/Ir(111) bilayers,
using local probes~\cite{Romming2013}. Overall, the realization of
skyrmionium metamatter should not be fundamentally more challenging than the
creation of other composite skyrmionic objects, such as skyrmion bags or
skyrmion bundles reported in chiral magnets including FeGe thin
films~\cite{tang2021magnetic}. A detailed discussion of experimental
protocols is, however, beyond the scope of the present theoretical work. The
goal of this study is to establish the physical principles and dynamical
properties of such states, which may stimulate future experimental
exploration.}

\subsection{The current-driven regimes}

In the present paper, we focus on the most salient aspects of the
\emph{current-driven} dynamics of Skm-based meta-matter, emphasizing the
qualitative changes that emerge under applied electric currents. In contrast
to isolated skyrmions, whose motion is primarily governed by a rigid-body
response to spin-transfer or spin--orbit torques, skyrmioniums and their
assemblies exhibit a much richer dynamical behavior stemming from their
internal structure and collective interactions.

For many of the dynamical processes discussed below, a faithful description
based solely on sequences of static snapshots is inherently limited.
The essential features of the transformations are more clearly revealed in
real-time evolution, which is best captured by supplementary videos \#5-9.
We therefore refer the reader to the accompanying videos, where the full
spatiotemporal development of the textures can be directly observed and
analyzed. As a representative point in the phase diagram, we choose $(0,0.31)$, which lies above the spiral-stability region and ensures that the quasi-atoms—skyrmioniums and skyrmions—retain a nearly circular shape.

Supplementary Video~\#5 illustrates the current-driven response of the
Skm$_2$Sk meta-matter at a current density of
$I = 1 \times 10^{12}\,\mathrm{A/m^{2}}$ ($\mathbf{j}||x$).
In this regime, the skyrmionium-based lattice responds elastically, preserving
its global topology and lattice symmetry while undergoing collective
translation accompanied by only weak internal deformations.
Skyrmioniums located along the trajectories of the faster-moving skyrmions
transiently contract and laterally displace, thereby opening a passage that
facilitates the unhindered motion of skyrmions through the lattice (Fig. \ref{fig10}(a)).

Supplementary Video~\#6 demonstrates that an applied current
($I = 1 \times 10^{12}\,\mathrm{A/m^{2}}$) can also induce a switching between
different Skm--Sk orders [Fig. \ref{fig10}(b)].
After an initial transient stage, the SkmSk polymorph with
$P2mm$ plane symmetry [Fig.~\ref{fig09}(c)] transforms into a polymorph
characterized by $P4gm$ symmetry [Fig.~\ref{fig09}(b)].
Following this symmetry reorganization, the composite lattice translates as a
rigid entity, with no relative displacements between the constituent
topological solitons.

The current-driven response of the SkmSk$_2$ meta-matter [Fig. \ref{fig09}(d)] strongly depends on the applied current density. At low currents ($I = 1 \times 10^{12}\,\mathrm{A/m^{2}}$; Supplementary Video~\#7), the lattice undergoes a brief transient deformation and subsequently
moves coherently as a rigid structure.
With increasing current density, additional dynamical channels become active.
At $I = 4 \times 10^{12}\,\mathrm{A/m^{2}}$ (Supplementary Video~\#8), skyrmions
perform hopping events between neighboring potential wells, leading to repeated
exchanges of their positions in a manner reminiscent of billiard-like
collisions [Fig. \ref{fig10}(c)].

Supplementary Movie~\#9 illustrates the current-driven dynamics of the
Skm$_2$Sk$_3$ meta-matter, revealing an intermediate response that bridges
collective lattice motion and soliton reorganization [Fig. \ref{fig10}(d)].
After a short transient period, the initially periodic structure undergoes a
self-organized rearrangement into quasi-one-dimensional vertical stripes
composed of alternating skyrmions and skyrmioniums.
In this dynamically generated stripe phase, skyrmioniums propagate along the
current direction while simultaneously exhibiting pronounced breathing
oscillations, reflecting their internal softness.
In contrast, skyrmions migrate predominantly along the stripe direction,
periodically replacing one another in a conveyor-like fashion.
This coexistence of longitudinal transport, transverse soliton exchange, and
internal mode excitation highlights the complex interplay between topology,
inter-soliton interactions, and current-induced forces in mixed Skm–Sk
meta-matter, and demonstrates how electric currents can dynamically reconfigure
both the geometry and the transport pathways of topological textures.

Such a laning transition provides a natural connection to recent work on driven
binary systems with mobility contrast \cite{vizarim2025}.
It has been shown that mixtures of particles with different effective mobilities
can spontaneously reorganize into phase-separated transport channels under
external driving, as observed for oppositely charged colloids in electric fields
and in pedestrian or active-matter systems where counterpropagating agents form
lanes that enhance collective mobility.
The stripe-like dynamical patterns observed in the Skm$_2$Sk$_3$ meta-matter
[Supplementary Video~\#9] represent a magnetic realization of this generic
phenomenon, with topology and the skyrmion Hall effect introducing an intrinsic
chirality that tilts the lanes relative to the driving direction.


Overall, the current-driven regimes of Skm meta-matter reveal a hierarchy of
dynamical responses---from elastic motion to internal-mode–dominated dynamics
and ultimately to structural instabilities---that have no direct analogue in
conventional skyrmion lattices. This richness reflects the interplay between
internal degrees of freedom, topology, and collective interactions, and forms
the basis for the unconventional transport phenomena. 
We emphasize that the examples discussed above represent only a limited
selection of possible current-driven responses of Skm-based meta-matter and are
by no means exhaustive.
Given the large parameter space and the diversity of accessible polymorphs, a
much more comprehensive and systematic analysis is required to fully classify
the dynamical regimes and transformation pathways.
Such an in-depth investigation, including a detailed mode analysis and a
broader exploration of lattice symmetries and compositions, lies beyond the
scope of the present work and will be addressed in future studies.

\section{Conclusions}

In this work, we have presented a comprehensive study of the current-driven
dynamics, transformations, and instability mechanisms of skyrmioniums in chiral
magnetic films, covering both isolated objects and collective states forming
skyrmionium-based meta-matter.
By combining  micromagnetic simulations with an analytical framework
based on the generalized Thiele equation, we have clarified how the internal
structure of skyrmioniums governs their nonequilibrium response to electric
currents.
{\color{black}In particular, the present study addresses several aspects of
skyrmionium dynamics that have remained largely unexplored, including the
microscopic origin of a finite skyrmionium Hall response, the instability
pathways of skyrmioniums under strong driving, and the collective transport
regimes emerging in skyrmionium-based meta-matter.}

For isolated skyrmioniums, we demonstrated that a finite transverse velocity—
the skyrmionium Hall effect—arises despite the vanishing total topological
charge.
We showed that this effect is  already present in static
skyrmioniums, originating from the imbalance between positive and negative
topological contributions associated with the inner skyrmion and the surrounding ring, which generally occupy different surface areas.
Upon current-induced deformation, this imbalance is enhanced, leading to Hall angles comparable to those of isolated skyrmions.
Our explicit evaluation of the gyrotropic and dissipative tensors provides a
quantitative link between micromagnetic simulations and the effective Thiele
description.
{\color{black}These results clarify the origin of the residual Hall response
sometimes observed in simulations and demonstrate that even weak discreteness
effects can break the perfect cancellation of topological contributions assumed
in ideal continuum descriptions.}

At higher driving currents, we identified several distinct instability channels
of isolated skyrmioniums.
Depending on magnetic field, anisotropy, and current density, skyrmioniums may
undergo unbounded elongation, collapse into an isolated skyrmion, transform into
a topologically trivial droplet, or expand into stripe-like textures that span
the system.
By tracking the temporal evolution of the topological charge and a local measure
of rotational intensity, we resolved the microscopic pathways of these
processes and clarified the role of singular and nonsingular spin-rotation
events.
{\color{black}Importantly, these instability mechanisms lie beyond the scope of
the Thiele equation, which assumes a rigid solitonic texture and therefore
cannot capture strong deformations of the magnetization field or transitions
between different magnetic states.}
We further showed that pulsed-current protocols introduce an additional control
parameter, allowing access to dynamic regimes that cannot be reached under
continuous driving while avoiding catastrophic annihilation events.

Beyond isolated textures, we explored the current-driven behavior of mixed
skyrmion–skyrmionium lattices, or Skm-based meta-matter.
We demonstrated that such composite lattices can respond elastically to weak
currents, undergo current-induced polymorphic transitions between states of
different plane-group symmetry, or exhibit strongly nonlinear dynamics
characterized by soliton exchange, confinement, and stripe formation.
{\color{black}These results extend the recently introduced concept of
skyrmionium meta-matter by revealing how interactions between topological
quasi-atoms with different charges and internal modes give rise to collective
transport phenomena that cannot be inferred from the dynamics of isolated
solitons.     

The phase diagram presented in this work is expressed in dimensionless units,
which allows the results to be applied to a broad class of material systems,
ranging from bulk chiral magnets with intrinsic
Dzyaloshinskii--Moriya interaction to multilayer films with
interfacially induced DMI. For a given material, once the microscopic
parameters---exchange stiffness, DMI constant, magnetic anisotropy, and
saturation magnetization---are known, the corresponding dimensionless
parameters $k_u$ and $h$ can be readily evaluated. This mapping makes it
possible to determine whether a particular material system falls within the
metastability region of skyrmioniums shown in the phase diagram.

As an example, the system PdFe/Ir(111), studied in our previous work
\cite{leonov2016properties}, is characterized by $k_u = 0.32$. According to the phase
diagram in Fig.~1 (e), the application of a magnetic field brings this material into the region where skyrmioniums become metastable. Moreover, depending on the field value, the system may enter regimes of skyrmionium instabilities illustrated in Figs.~4 and~5. In particular, below the line $D--B$ skyrmioniums exist within the stability region of the skyrmion lattice, whereas above this line they occur in the region where the ferromagnetic state represents the thermodynamically stable phase.}

Overall, our findings show that current-induced degradation processes of
skyrmioniums are not merely failure modes but instead provide direct insight into
the underlying topological energy landscape of chiral magnets far from
equilibrium.
{\color{black}More broadly, the present results help bridge the gap between
idealized continuum descriptions and realistic discrete-spin systems,
highlighting the role of internal structure and lattice effects in determining
skyrmionium dynamics.}
The rich hierarchy of dynamical regimes uncovered here underscores the potential
of skyrmioniums and their meta-matter assemblies as tunable building blocks for
future spintronic and magnonic applications, while also motivating further
systematic studies of their collective dynamics beyond the examples considered
in the present work.

\section{Acknowledgments}
A.O.L. acknowledges JSPS Grant-in-Aid (C) 26K06945.

\section{Supplementary Material}


The current-driven dynamics and transformation pathways of isolated skyrmioniums
(Supplementary Videos~1--4) and skyrmionium-based meta-matter
(Supplementary Videos~5--9) are inherently time dependent and involve collective
motion, internal deformation, and topological rearrangements that cannot be
fully captured by static figures alone. To complement the results presented in
the main text, we therefore provide a set of Supplementary Videos visualizing the
real-time evolution of the magnetization textures.

Unless stated otherwise, the videos display color maps of the out-of-plane
magnetization component $m_z$, with arrows indicating the in-plane magnetization
direction. In selected cases, the in-plane component $m_x$ or the topological
charge density is shown explicitly, as specified in the corresponding captions.
For isolated skyrmioniums, a comoving reference frame is adopted, such that the
soliton appears stationary and its internal dynamics and instability mechanisms
can be clearly resolved. In some videos, the view is additionally zoomed into
regions where the transformations are initiated.

The magnetization components are visualized using the red--white--blue color
scheme native to \textsc{MuMax3}, while the topological charge density is encoded
using the \texttt{turbo} colormap in \textsc{Matlab}.

\medskip
\noindent\textbf{Supplementary Video~1:}
Instantaneous collapse of a skyrmionium into an isolated skyrmion
(region~II in Fig.~4(a)), occurring via the formation of defects.
The video successively shows the distributions of the topological charge density
and the $m_z$ and $m_x$ magnetization components during the collapse process.

\medskip
\noindent\textbf{Supplementary Video~2:}
Formation of a meron-tipped stripe protruding from the skyrmionium boundary and
undergoing elongation (region~III in Fig.~5(a)).
The first half of the video shows the evolution of the out-of-plane magnetization
component $m_z$, while the second half displays the corresponding topological
charge density.

\medskip
\noindent\textbf{Supplementary Video~3:}
Collapse of a skyrmionium into an isolated skyrmion (region~II in Fig.~5(a)),
proceeding via a mechanism distinct from that shown in Supplementary Video~1.
The central skyrmion becomes progressively exposed to the surrounding
ferromagnetic background from the right-hand side and ultimately collapses.

\medskip
\noindent\textbf{Supplementary Video~4:}
Instability of a skyrmionium with respect to the formation of a topologically
trivial droplet (region~IV in Fig.~5(a)).
The transformation proceeds through the same initial stages as in Supplementary
Video~3; however, instead of the mutual annihilation of the pair with opposite
winding numbers, only the component carrying positive winding number $q$ is lost
due to its gradual connection to the homogeneous ferromagnetic background. The
remaining component with negative winding number persists and forms the upper
tip of the emerging droplet.

\medskip
\noindent\textbf{Supplementary Video~5:}
Elastic current-driven motion of Skm$_2$Sk meta-matter at
$I = 1 \times 10^{12}\,\mathrm{A/m^{2}}$.
The composite lattice undergoes collective translation while preserving its
overall topology and plane-group symmetry. Only weak internal deformations occur:
skyrmioniums located along the paths of faster-moving skyrmions transiently
contract and shift laterally, thereby opening corridors that facilitate
skyrmion transport through the lattice.

\medskip
\noindent\textbf{Supplementary Video~6:}
Current-induced switching between Skm--Sk polymorphs at
$I = 1 \times 10^{12}\,\mathrm{A/m^{2}}$.
After an initial transient, the SkmSk lattice with $P2mm$ symmetry transforms
into a configuration with $P4gm$ symmetry. Following this symmetry
reorganization, the entire meta-matter translates as a rigid body without
relative motion between individual skyrmions and skyrmioniums.

\medskip
\noindent\textbf{Supplementary Video~7:}
Low-current dynamics of SkmSk$_2$ meta-matter
($I = 1 \times 10^{12}\,\mathrm{A/m^{2}}$).
After a brief deformation stage, the lattice recovers its integrity and moves
coherently as a single object, indicating a predominantly elastic response of
the composite structure.

\medskip
\noindent\textbf{Supplementary Video~8:}
High-current dynamics of SkmSk$_2$ meta-matter at
$I = 4 \times 10^{12}\,\mathrm{A/m^{2}}$.
Additional dynamical channels become activated, leading to hopping events of
skyrmions between neighboring potential wells. These jumps result in repeated
position exchanges between skyrmions in a billiard-like manner, while
skyrmioniums act as a dynamically reconfigurable background.

\medskip
\noindent\textbf{Supplementary Video~9:}
Current-driven reorganization of Skm$_2$Sk$_3$ meta-matter at
$I = 1 \times 10^{12}\,\mathrm{A/m^{2}}$.
After a transient period, the initially periodic lattice self-organizes into
quasi-one-dimensional stripes composed of alternating skyrmions and
skyrmioniums. In this stripe phase, skyrmioniums propagate predominantly along
the driving direction while exhibiting pronounced breathing oscillations,
whereas skyrmions migrate along the stripe direction and periodically replace
one another. This laning-type dynamics highlights the coexistence of collective
transport, internal-mode excitation, and topology-driven soliton exchange in
mixed Skm--Sk meta-matter.

\bibliographystyle{apsrev4-2}
\bibliography{refs}

\end{document}